\documentclass[twocolumn]{aastex62}
\usepackage{apjfonts}

\newcommand{\cajwl}{$Ca_{\rm JWL}$}
\newcommand{\nhjwl}{$nh_{\rm JWL}$}
\newcommand{\chjwl}{$ch_{\rm JWL}$}
\newcommand{\cnjwl}{$cn_{\rm JWL}$}
\newcommand{\cubi}{$C_{\rm UBI}$}
\newcommand{\cnw}{CN-w}
\newcommand{\cns}{CN-s}
\newcommand{\dvbump}{$\Delta V_{\rm bump}$}

\newcommand{\hkjwl}{$hk_{\rm JWL}$}
\newcommand{\hst}{{\it HST}}

\newcommand{\nrgb}{$n$(\cnw):$n$(\cns)}
\newcommand{\nsub}{$n$(SP1):$n$(SP2)}

\newcommand{\pchjwl}{$\parallel ch_{\rm JWL}$}
\newcommand{\pcnjwl}{$\parallel cn_{\rm JWL}$}
\newcommand{\pnhjwl}{$\parallel nh_{\rm JWL}$}

\newcommand{\phkjwl}{$\parallel hk_{\rm JWL}$}

\newcommand{\str}{Str\"omgren}
\newcommand{\vbump}{$V_{\rm bump}$}

\newcommand{\vvhb}{$V - V_{\rm HB}$}
\newcommand{\vvhbmag}{$-$2 mag $\leq$ $V - V_{\rm HB}$ $\leq$ 2 mag}
\newcommand{\cnwave}{$\lambda$3883}
\newcommand{\chwave}{$\lambda$4250}
\newcommand{\nhwave}{$\lambda$3360}
\newcommand{\dy}{$\Delta$Y}
\newcommand{\dc}{$\Delta_{C~{\rm F275W,F814W}}$}
\newcommand{\wfg}{$W^{1G}_{F275W-F814W}$}
\newcommand{\dtrio}{$\Delta_{C~{\rm F275W,F336W,F438W}}$}

\newcommand{\gaia}{{\it Gaia}}
\newcommand{\ebv}{$E(B-V)$}
\newcommand{\eby}{$E(b-y)$}
\newcommand{\ehk}{$E$(\hkjwl)}
\newcommand{\caii}{\ion{Ca}{2}}
\newcommand{\nnnc}{$n$(N)/$n$(C)}
\newcommand{\agbm}{AGB-manqu\'e}
\newcommand{\kms}{km sec$^{-1}$}
\newcommand{\ciso}{$^{12}$C/$^{13}$C}

\newcommand{\fehhk}{[Fe/H]$_{hk}$}
\newcommand{\cfecn}{[C/Fe]$_{\rm CN}$}
\newcommand{\nfecn}{[N/Fe]$_{\rm CN}$}
\newcommand{\msun}{$M_{\rm \odot}$}

\shorttitle{M3}
\shortauthors{Lee \& Sneden}
\begin{document}
 
\title{Multiple Stellar Populations of Globular Clusters from Homogeneous Ca-CN-CH-NH Photometry. VI. M3 (NGC 5272) is not a Prototypical Normal Globular Cluster
\footnote{Based on observations made with the Kitt Peak National Observatory (KPNO) 0.9 m telescope, which is operated by WIYN Inc. on behalf of a Consortium of partner Universities and Organizations.}
\footnote{This work has made use of data from the European Space Agency (ESA) mission \gaia\ (\url{https://www.cosmos.esa.int/gaia}), processed by the \gaia\ Data Processing and Analysis Consortium (DPAC, \url{https://www.cosmos.esa.int/web/gaia/dpac/consortium}). Funding for the DPAC has been provided by national institutions, in particular the institutions participating in the \gaia\ Multilateral Agreement.}
}

\author[0000-0002-2122-3030]{Jae-Woo Lee}
\affiliation{Department of Physics and Astronomy, Sejong University, 209 Neungdong-ro, Gwangjin-Gu, Seoul, 05006, Korea, jaewoolee@sejong.ac.kr, jaewoolee@sejong.edu}
\affiliation{
Department of Astronomy and McDonald Observatory, The University of Texas, Austin, TX 78712, USA}
\author{Christopher Sneden}
\affiliation{
Department of Astronomy and McDonald Observatory, The University of Texas, Austin, TX 78712, USA}

\begin{abstract}
We present Ca-CN-CH-NH photometry for the well-known globular cluster (GC) M3 (NGC 5272). We show new evidence for two M3 populations with distinctly different carbon and nitrogen abundances, seen in a sharp division between CN-weak and CN-strong red-giant branches (RGBs) in M3. The CN-strong population shows a C-N anticorrelation that is a natural consequence of the CN cycle, while the CN-weak population shows at most a very weak C-N anticorrelation. Additionally, the CN-weak population exhibits an elongated spatial distribution that is likely linked to its fast rotation. Our derived metallicities reveal bimodal distributions in both populations, with $\langle$[Fe/H]$\rangle\approx-$1.60 and $-$1.45, which appear to be responsible for the discrete double RGB bumps in the CN-weak and the large $W^{1G}_{F275W-F814W}$ range. From this discovery, we propose that M3 consists of two GCs, namely C1 (23\%, $\langle$[Fe/H]$\rangle\approx-1.60$) C2 (77\%, $\langle$[Fe/H]$\rangle\approx-1.45$), each of which has its own C-N anticorrelation and structural and kinematical properties, which are strong indications of independent systems in M3. The fractions of the CN-weak population for both the C1 and C2 are high compared to Galactic GCs but they are in good agreement with GCs in the Magellanic Clouds.  We suggest that M3 is a merger remnant of two GCs, most likely in a dwarf galaxy environment, and accreted to our Galaxy later in time. This is consistent with recent proposals of an ex-situ origin of M3.
\end{abstract}


\section{Introduction}
Galactic globular clusters (GCs) have been extensively studied to understand the formation and evolution of our Galaxy. 
For several decades the existence of not only the in-situ but also the accreted or ex-situ GCs has been recognized \citep{sz,zinn85,zinn93,lc99b} and interpreted in terms of cold dark matter cosmology that predicts a hierarchical structure formation in the universe.
The \gaia\ satellite has radically improved the situation with astrometry and proper motion data for more than a billion of stars \citep{gaiadr2}. 
Previously challenging tasks of identifying merger events in our Galaxy can now be realized and several remnants of the ancient merger events have been proposed \citep[e.g., see][]{helmi18,myeong19}.

It has been thought that younger halo GCs, including M3, were likely accreted by our Galaxy in the past to form  components of the Galactic halo seen today \citep[e.g.,][]{zinn93,lc99b}.
For example, \citet{kruijssen19} suggested that M3 is likely an ex-situ GC and a member of a hypothetical galaxy, the so-called {\it Kraken}, that accreted into our Galaxy.
More recently, \citet{koppelman19} argued that M3 belongs to the {\it Helmi Stream} that is likely accreted about 5--8 Gyr ago, in which seven GCs exhibit a tight age-metallicity relation.

M3 has long been considered as a prototypical ``normal'' GC. 
But M3 lies relatively far from the Galactic center, and it has some interesting structural aspects.
It is one of the most massive GCs in our Galaxy and, most interestingly, it is the most RR Lyrae (RRL) rich GC in our Galaxy, possessing more than 240 RRLs, which makes M3 exceptional \citep[][the 2019 March version\footnote{See C.~M.\ Clement, Catalogue of GC variable stars, available at \url{http://www.astro.utoronto.ca/~cclement/cat/listngc.html}).}]{clement01}.
A second RRL-rich GC is $\omega$ Cen, which is most likely the remnant of the core of the dwarf galaxy that accreted into our Galaxy and about 3.5 times more massive than M3 \citep{kruijssen19}. $\omega$ Cen has about 200 RRLs but it contains a significant fraction of metal-poor stars that do not pass through the instability strip during their core helium phase. A third RRL-rich GC is M5, which has similar mass as M3 but has about 130 RRLs.

M3 has an abnormally large range of the \hst\ pseudo-color index, \dc, of its first generation (FG) stars. 
This has led to investigations of amount of helium spread in the M3 FG \citep{lardo18,tailo19}, which concluded that it cannot explain the \dc\ spread within our current understanding of the GC formation and chemical evolution scenarios.
In our previous study of M3 CN band strengths  \citep{lee19a}, 
we found that the M3 \cnw\ population has an extended and tilted red-giant branch bump (RGBB), which later we will link to the large range of the \dc\ index.

Light elemental abundance variations in GCs were first identified about 50 years ago, and observational evidence for their ubiquitous nature has steadily increased \citep[e.g.,][]{osborn71,cohen78,norris81,sneden92}.
Explanations have centered on the idea of multiple stellar formation episodes, where the later generation of stars formed out of interstellar media polluted from the previous generation of stars \citep[e.g.][]{dercole08}.
During the past decade, a great deal of observational evidence of multiple populations (MPs) in GCs in the Milky Way has been emerged \citep[e.g.,][]{carretta09,lee09b,milone17,gratton19,marino19}. 
In spite of much effort, understanding the formation of GCs with MPs is not yet solved.
So far, none of the proposed models can explain the observational constraints of GC MPs with satisfaction \citep[e.g.,][]{bastian18}. 
Furthermore, MPs exist in GCs of external low-mass local-group galaxies, such as the Magellanic clouds \citep[e.g.,][]{mucciarelli09,milone20}, and the Fornax dwarf galaxy \citep[e.g.,][]{letarte06}, which poses another formidable problem of environmental effects on the formation of GCs with MPs. 
Revolutionary spectroscopic and photometric studies of MPs in extragalactic GCs will come with the advent of 30 m class telescope within the next decade. 
Until then, the best way to understand environmental effects on GC formation is to investigate the accreted or ex-situ GCs that are located near us.

In this paper, we investigate M3 using our own photometric system optimized for measuring carbon and nitrogen abundances, which are key elements for GC MP studies. 
In spite of the importance of the carbon and nitrogen abundance in the evolution of metal-poor low-mass stars, their abundances in GCs are often poorly known due to the lack of measurable atomic absorption lines in the visual or infrared wavelength regime. 
Instead, their elemental abundances can be measured via diatomic molecules, such as NH, CN, CH, and CO.

Studies of C and N abundances from visual or infrared high-resolution spectroscopy have been restricted to very bright red-giant branch (RGB) or asymptotic giant branch (AGB) stars, where the surface C and N abundances can be significantly altered from their primordial values due to the onset of CN-cycle accompanied by a non-canonical thermohaline mixing \citep{charbonnel07}. 
Low-resolution spectroscopy has been frequently applied to faint GC stars but traditional spectroscopy cannot be used in very crowded regions, such as the central part of GCs, due to stellar blending. 
Even for the isolated stars, low-resolution spectroscopy is also vulnerable to the selection of the continuum sidebands in the wavelength region of numerous strong absorption lines, which can compromise the measured spectrum indices \citep[e.g., see][]{lee19a,lee19c}.
Our approach mitigates both of these problems, through careful feature and passband choices for NH, CN, and CH photometric indices to provide reliable carbon and nitrogen abundances, and avoidance of stellar targets with potential contamination by other cluster members.

The work presented here will provide observational evidence that M3 is most likely a merger remnant of two GCs in a dwarf galaxy environment, where the relative velocity of the two GCs is smaller than their velocity dispersion \citep[e.g.,][]{gavagnin16}. 
The outline of this paper is as follows.
We describe our new filter system, $JWL34$, which is designed to measure the absorption strength of the NH bands at \nhwave\ in \S\ref{s:jwl34}. 
We present our observations and data reductions in \S\ref{s:obs}. 
The definitions of our color indices, including the new \nhjwl, will be given in \S\ref{s:tag} along with a discussion of our new strategy of populational tagging of M3 RGB stars.
In \S\ref{s:hst}, we show comparisons of our photometric indices with widely used color indices from the \hst\ photometry \citep{milone15}. 
We will discuss the metallicity dependency on the \dc\ index, which is the key to understand the large \wfg\  range of the M3 FG stars that has been pointed out in previous studies  \citep{lardo18,tailo19}.
In \S\ref{s:abund} we discuss synthetic model grids generated to interpret our observed indices into [Fe/H],\footnote{We adopt the standard spectroscopic notation \citep{wallerstein59} that for elements A and B, [A/B] $\equiv$ log$_{\rm 10}$(N$_{\rm A}$/N$_{\rm B}$)$_{\star}$ $-$ log$_{\rm 10}$(N$_{\rm A}$/N$_{\rm B}$)$_{\odot}$.
We equate metallicity with the stellar [Fe/H] value.}
[C/Fe] and [N/Fe] abundances. In particular, we will discuss the bimodal metallicity distributions for both the \cnw\ and \cns, which are the important clue that requires a new formation scenario for M3. We also discuss the separate C-N correlations between the two populations. Populational tagging for red horizontal branch (RHB) stars will be presented in \S\ref{s:rhb}, where we will show that populational number ratio of RHB stars is significantly different but can be naturally understood from the evolution of the metal-poor low-mass stars.
The structural and kinematical differences between MPs will be discussed in \S\ref{s:struct}. The discussion on the discrete double RGBBs of the \cnw\ will be given in \S\ref{s:rgbb},  where we will argue that the bimodal metallicity distribution is essential to explain the double RGBBs in the \cnw\ and the large extent of the \wfg\ of the M3 FG obtained with \hst\ photometry.
In \S\ref{s:merger}, we show that M3 is a merger remnant of two distinct GCs, having different elemental abundances, structural and kinematical properties.
Finally, our summary of the work will be provided in \S\ref{s:sum}.

\begin{figure}
\epsscale{1.1}
\figurenum{1}
\plotone{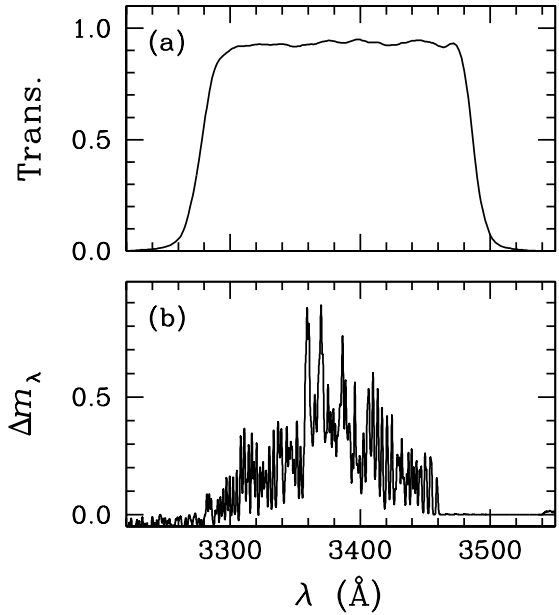}
\caption{
(a) The filter transmission function of our new $JWL34$ filter.
(b) Differences in monochromatic magnitude between the \cnw\ and \cns\ using the synthetic spectra for an arbitrary intermediate metallicity ([Fe/H] = $-$1.5) RGB stars at the level of the RGBB.
}\label{fig:jwl34}
\end{figure}

\section{$JWL34$: A New NH Filter System}\label{s:jwl34}
Our \cnjwl\ and \chjwl\ indices (discussed in 
Section~\ref{s:tag}) can provide significant information on
MPs in GCs  \citep[e.g.,][]{lee17, lee18, lee19a, lee19c, lee20}. 
Our \cnjwl\ index is satisfactory in most cases but there are three issues  that complicate attempts to transform \cnjwl\ indices into unambiguous nitrogen abundances.

First, the formation of CN molecules depends on both the carbon and nitrogen abundances. In low-luminosity less evolved GC stars with abundance ratios \nnnc\ $<$ 1, CN band strengths are mainly dependent on nitrogen abundances \citep[e.g.,][]{suntzeff81, briley93}. 
But this abundance ratio condition is not always true. For second generation (SG) stars in GCs with abundance ratios \nnnc\ $>$ 1, CN band strengths no longer scale monotonically with nitrogen abundance and can even decrease with depletion in the surface carbon abundances \citep[e.g.,][]{smith86}.

Second, CN is a double-metal diatomic molecule, and its band absorption strengths rapidly decline with decreasing metallicity \citep[e.g.,][]{sneden74,langer92}. On the other hand, absorption strengths of single-metal diatomic molecules (e.g., CH and NH) weaken only slowly with decreasing metallicity, They are still sensitive to variations in the carbon and nitrogen abundances in very metal-poor stars.

Third, \cfecn\ and \nfecn\ abundances derived from CN bands at \cnwave\ with prior [N/Fe] and [C/Fe] information supplied from \nhjwl\ and \chjwl\ indices depend somewhat on [O/Fe] abundances and \ciso\ ratios.
In particular, the oxygen abundances of individual stars within a given population have a substantial spread \citep[e.g., see][]{carretta09, marino19}, which can result in non-negligible differences in \cfecn\ and \nfecn\ in the absence of [O/Fe] abundances of individual stars. 
As shown in Appendix A, at metallicity of [Fe/H] = $-$1.5 dex, $\Delta$[O/Fe] = +0.1 dex can result in overestimation of \nfecn\ by about 0.08 dex (see Figure~\ref{fig:ap:ofe}). 
At the bright RGB (bRGB),\footnote{The bRGB stars and the faint RGB (fRGB) stars indicate stars brighter than and fainter than the RGBB, respectively.} the \nfecn\ is also vulnerable to uncertain \ciso\ ratios, which are often undetermined for GC RGB stars. 
The difference in \nfecn\ with the \ciso\ = 5 and 90 ($\approx$ a solar carbon isotope ratio) can be as large as 0.5 dex (see Figure~\ref{fig:ap:ciso}). 

In an attempt to directly measure the NH band at \nhwave, we have developed a new filter, $JWL34$. In panel (a) of Figure~\ref{fig:jwl34} we show the transmission function of our $JWL34$ filter, which has a pivot wavelength of $\approx$ $\lambda$337 nm and a full-width at the half-maximum of $\Delta\lambda$ $\approx$ 20.5 nm.
In panel (b) of this figure we show the sensitivity of $JWL34$ to N abundance changes by plotting the difference in monochromatic magnitudes between two intermediate-metallicity ([Fe/H]~$-$1.50) stars: a \cnw\ star (with assumed abundances [C/Fe]~=~$-$0.10, [N/Fe]~=~0.10) and a \cns\ star ([C/Fe]~=~$-$0.30, [N/Fe] =~0.80). 
The monochromatic magnitudes are those that were obtained with synthetic spectrum computations, as will be described in detail in Section~\ref{ss:CN}.

\begin{figure}
\epsscale{1.2}
\figurenum{2}
\plotone{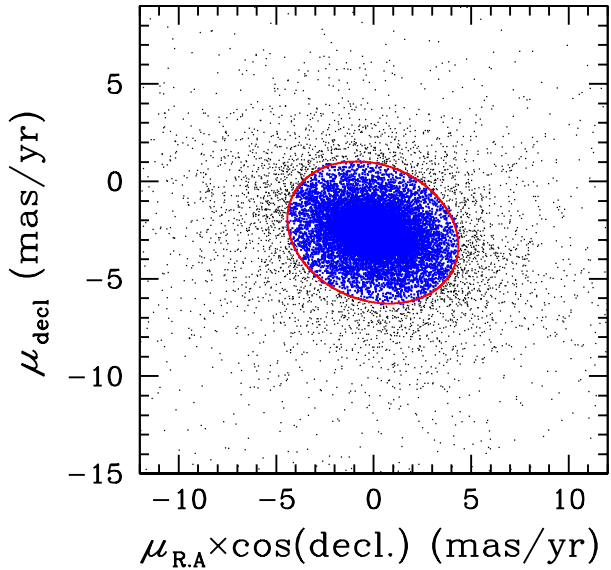}
\caption{
\gaia\ DR2 proper motions of our FOV. The red ellipse indicates the boundary (3$\sigma$) of the M3 member stars shown with blue dots.
}\label{fig:pm}
\end{figure}

\begin{deluxetable*}{cccccccccccc}
\tablenum{1}
\tablecaption{Integration times for M3 (s)\label{tab:obs}}
\tablewidth{0pc}
\tablehead{
\colhead{} &
\colhead{$y$} & \colhead{$b$} & \colhead{$Ca_{\rm JWL}$} & \colhead{$JWL39$} & 
\colhead{$JWL43$} & \colhead{$JWL34$} &
\colhead{} & \colhead{$V$} & \colhead{$B$} & \colhead{$I$} & \colhead{$U$}
}
\startdata
NGC~5272(M3)    & 8,910 & 19,520 & 52,000 & 23,500 & 24,000 & 19,400 & & 1,380 & 3,070 & 1,220 & 11,350 \\
\enddata 
\end{deluxetable*}

\begin{figure*}
\epsscale{1}
\figurenum{3}
\plotone{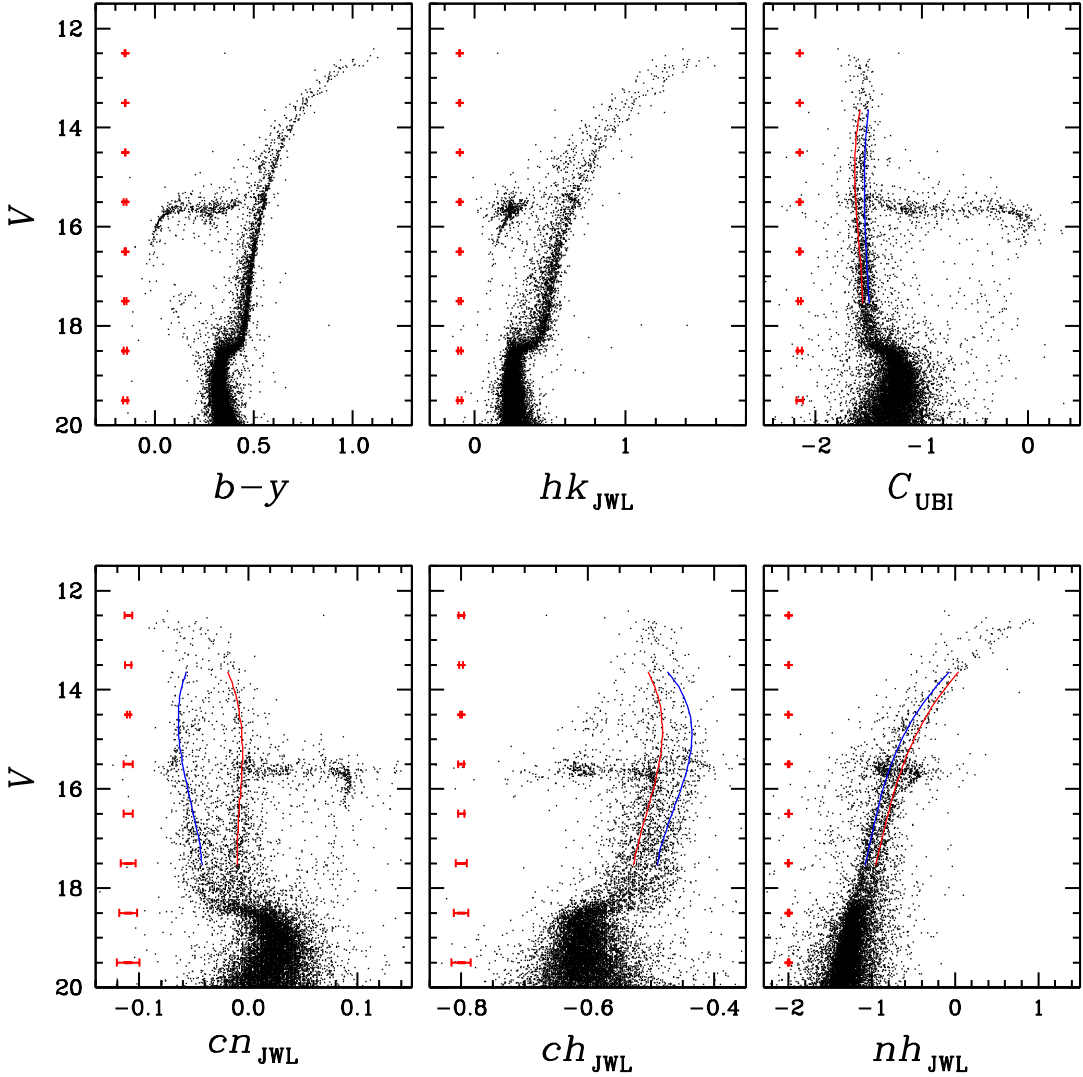}
\caption{
CMDs with various color axes of M3 member stars based on the proper motion study of the \gaia\ DR2. The red error bars represent the mean photometric measurement uncertainties at given magnitude bins. The fiducial sequences to calculate the parallelized color indices are also shown.
}\label{fig:cmd}
\end{figure*}

\begin{figure}
\epsscale{1.2}
\figurenum{4}
\plotone{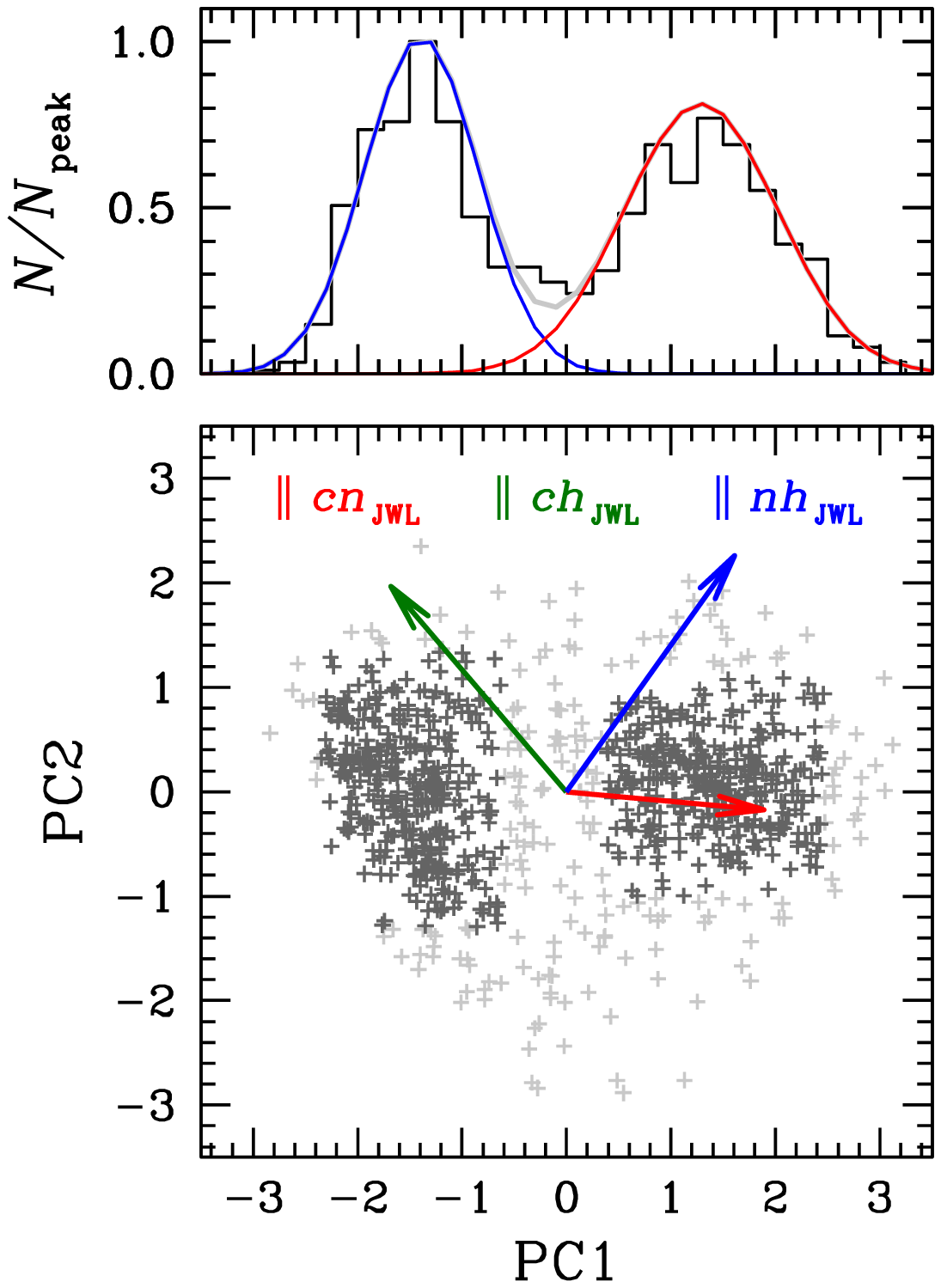}
\caption{
(Bottom panel) A plot of PC1 versus PC2, calculated using \pcnjwl, \pchjwl, and \pnhjwl. Individual eigenvectors are also shown.
All stars are shown with light gray plus signs, while dark gray plus indicate stars within 2$\sigma$ from the center of each population in each axis.
(Top panel) Populational tagging using the expectation maximization method along the PC1.
}\label{fig:pca}
\end{figure}

\section{Photometric Data}\label{s:obs}
In 2017 and 2018, we obtained photometric data for M3 in 20 nights in 5 separate runs using the Half Degree Imager (HDI), which is equipped with an e2V 4k $\times$ 4k CCD chip, mounted on the the KPNO 0.9m telescope. 
The HDI provides a field of view (FOV) of 30\arcmin $\times$ 30\arcmin. 
The total integration times of \str\ $y$, $b$, \cajwl, $JWL39$, and $JWL43$ for the M3 science field are given in Table~1 \citep[see also][]{lee19a}.

Additional photometric data for M3 using our new $JWL34$ filter and \str\ $y$ and $b$ filters were collected in six nights from 2019 June 27 to July 5. 
Integration times for $JWL34$, \str\ $y$, and $b$ filters were 19,400 s, 1,220 s, and 2,200 s. 
Due to the electronic problem with the HDI in 2019, we used the S2KB CCD cam, which is equipped with a 2k $\times$ 2k CCD chip and provides a FOV of 21\arcmin $\times$ 21\arcmin.

Our integration times for individual filters are long enough to perform accurate photometry of RGB stars in M3. Typical photometric measurement errors at the level of the RGBB in individual color indices are less than 0.01 mag, so the broad or bimodal RGB sequences in some color indices are real features in M3, not artifacts 
arising from measurement errors.

Limitations in spatial angular resolving power of small aperture ground-based telescopes, such as the KPNO 0.9m telescope, lead to incomplete detection and large photometric measurement uncertainties in central parts of GCs. 
Deriving accurate populational number ratios requires detection of statistically robust samples with reliable measurements in the central part of GCs showing strong radial populational gradient such as M3, but this is not an important point in our current study. 
Most of the analyses presented in this paper will be based on stars with very accurate measurements located $r \geq$ 1\arcmin, and we did not apply the method that we developed for using prior positional information from \hst\ observations \citep[e.g., see Appendix B of][]{lee17}. 
As we will discuss below, we made use of the proper motion study from the second \gaia\ data release \citep{gaiadr2}, in which the source detection incompleteness becomes large in the central part that also hinders our populational tagging of the central part of the cluster.

The interstellar reddening of M3 is very small, \ebv\ = 0.01 \citep[][the 2010 edition]{harris96}\footnote{Available at \url{http://physwww.mcmaster.ca/~harris/mwgc.dat}.}. 
Therefore, any differential reddening across the science field will be undetectably small and will not affect our results (see also \S\ref{ss:rgbb_red}). 

Reductions of our raw photometric data have been discussed in detail by \cite{lee15} and will not be repeated here.
The photometry of M3 and standard stars were analyzed using DAOPHOTII, DAOGROW, ALLSTAR and ALLFRAME, and  COLLECT-CCDAVE-NEWTRIAL packages \citep{pbs87,pbs94,lc99a}. 
We derived astrometric solutions for individual stars using the data extracted from the Naval  Observatory Merged Astrometric Dataset \citep[NOMAD,][]{nomad} and the IRAF IMCOORDS package.  

We made use of the proper motions from the second \gaia\ date release to select the cluster's membership stars \citep{gaiadr2}, following the method similar to those used in our previous studies, \citep[see, e.g.,][and references therein]{lee20}.
We derived the mean values of proper motions of M3 with iterative sigma-clipping calculations, finding that, in an unit of mas/yr, ($\mu_{\rm RA}\times\cos\delta$, $\mu_{\rm decl}$) = ($-$0.039, $-$2.622) with standard deviations along the major axis of the ellipse of 1.509 mas/yr and along the minor axis of 1.155 mas/yr. 
We considered that stars within 3$\sigma$ from the mean values to be M3 member stars as shown in Figure~\ref{fig:pm}. Then we selected our target RGB stars with \vvhbmag\ from our multi-color photometry. 

The Galactic latitude of the cluster is high, 79\degr. Therefore the contamination by the off-cluster field stars will not be severe in our photometry. For example, we estimate that the total number of field stars with the magnitude range of our interest (\vvhbmag) toward our M3 science FOV would expect to be $\approx$ 140 \citep{ratnatunga85}, which is corresponding to only 6.4 $\pm$ 0.6 \% of our total number of proper motion membership stars in the magnitude range of our interest.
In addition to proper motion study, as  \citet{lee15,lee20} demonstrated, our \cajwl\ photometry provides a powerful means to distinguish between the GC RGB stars with low metallicities and Galactic disk stars with high metallicities.
Therefore we believe that the contribution from off-cluster field stars is too small to affect our results presented here.

\begin{figure*}
\epsscale{1}
\figurenum{5}
\plotone{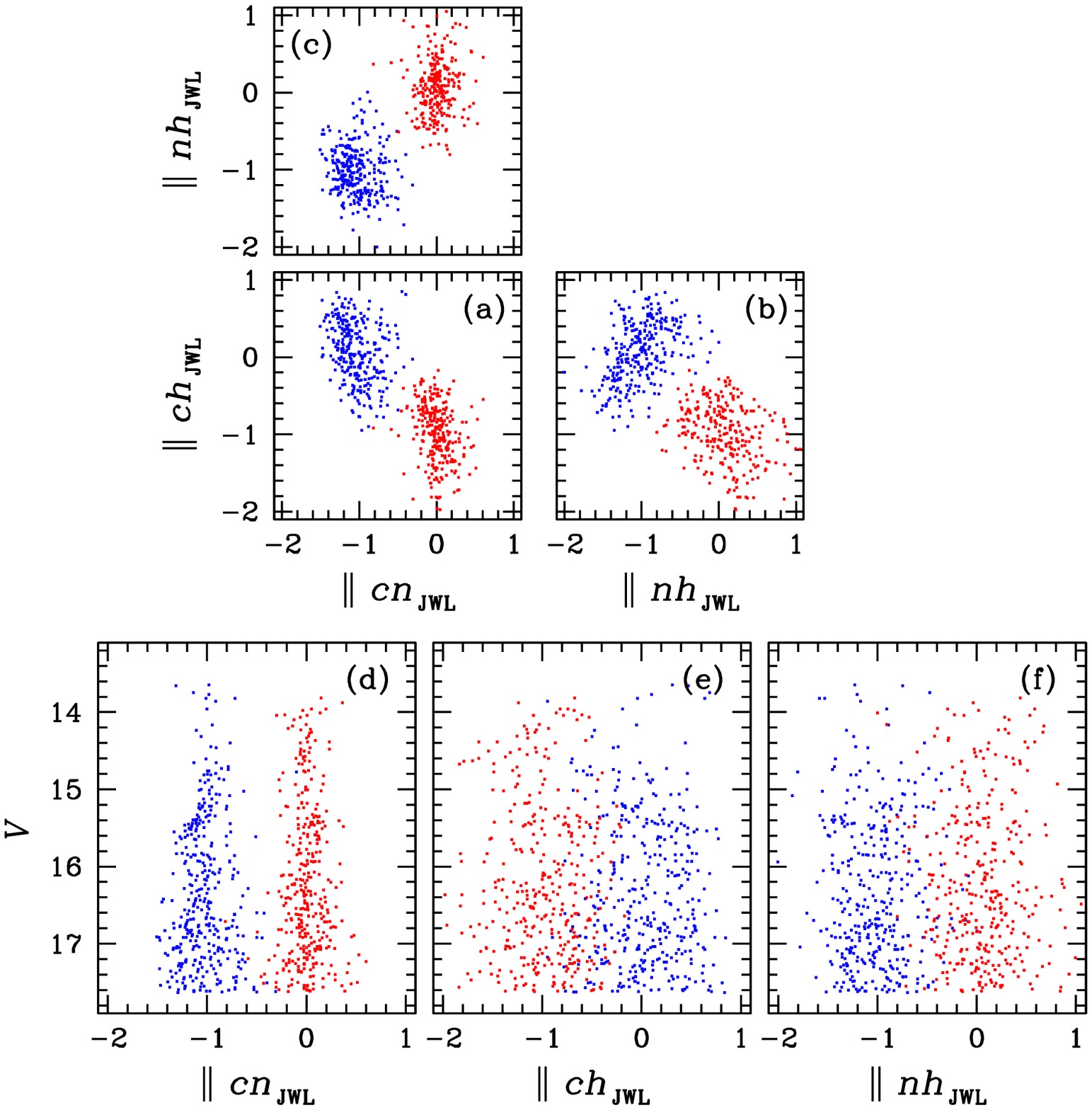}
\caption{
(a) The photometric \pcnjwl\ versus \pchjwl\ relation for RGB stars in the 
magnitude range \vvhbmag\ in M3. 
(b) The \pnhjwl\ versus \pchjwl\ relation.
(c) The \pcnjwl\ versus \pnhjwl\ relation.
(d) The \pcnjwl\ versus $V$ CMD. 
(e) The \pchjwl\ versus $V$ CMD.
(f) The \pnhjwl\ versus $V$ CMD.
In all panels, the blue dots denote the \cnw\ population while the red dots 
denote the \cns\ population.
}\label{fig:cnchnh}
\end{figure*}

\section{Populational Tagging: New Strategy}\label{s:tag}
Throughout this work, we will use our own photometric indices, defined as,
\begin{eqnarray}
hk_{\rm JWL} &=& (Ca_{\rm JWL} - b) - (b-y). \label{eq:hk} \\
nh_{\rm JWL} &=& (JWL34 - b) - (b-y). \label{eq:nh} \\
cn_{\rm JWL} &=& JWL39 - Ca_{\rm JWL}, \label{eq:cn} \\
ch_{\rm JWL} &=& (JWL43 - b) - (b-y). \label{eq:ch} 
\end{eqnarray}
These relationships have functional forms analogous to traditional color indices, and will be treated as such throughout this paper.

Assuming constant [Ca/Fe] ratios \citep[e.g., see][]{carney96, marino19}, the \hkjwl\ index is a good photometric measure of metallicity, with a weak CH band contamination \citep{att91, lee09a, lee09b, lee15, lee19c}. 
In Appendix~C and Figure~\ref{fig:ap:feh}, we show the \hkjwl\ dependency on metallicity \citep[see also Supplementary Information of][]{lee09a}. Also in Appendix~D and Figure~\ref{fig:ap:CFe2CaFe}, we show the \hkjwl\ dependency on carbon abundance, which is almost nil for M3.

As we discussed in our previous works, \cnjwl\ and \chjwl\ are excellent photometric measures of the CN band at \cnwave\ and CH G band at \chwave, respectively, for cool stars \citep[see][and references therein]{lee20}. 

In our study of M3, the \cnjwl\ index is still powerful in classifying MPs, as we will show below. For cool GC RGB stars, the emergent stellar surface flux near \cnwave\ is much greater than that near \nhwave. At the same time, the degree of interstellar and atmospheric extinctions for $JWL39$ is much smaller than those for $JWL34$ \citep[e.g., see Figure 2 of][]{lee17}. Therefore, more accurate photometry can be attained with \cnjwl\ than with \nhjwl.
In addition, populational tagging from the \cnjwl\ can be more readily accomplished than by using \chjwl\ or \nhjwl\ indices, as long as the C-N anticorrelation holds.
For example, the CH distributions of GC RGB stars may not show clear bimodality (e.g., see Figure~6 of \citealt{norris84}). 
But the CN bimodality in GCs has been known for decades and frequently studied for GC stars \citep[e.g.,][]{osborn71}.

In Figure~\ref{fig:cmd}, we show CMDs of M3 member stars based on the proper motion study of \gaia\ DR2 \citep{gaiadr2}. 
We also show measurement uncertainties for each color index, suggesting that the discrete double RGB sequences or the spread in color indices in the magnitude range of our interest (\vvhbmag) are real features; they are much larger than photometric errors.
Our \cnjwl\ CMD (bottom left panel) shows conspicuous double RGB sequences in M3; these have been known for decades \citep[e.g,][]{suntzeff81,briley93}.
Our \chjwl, \nhjwl, and \cubi\ CMDs also show weak bimodalities or large spread in their color indices, suggestive of heterogeneous CNO abundances among RGB stars.

In order to remove the luminosity effect on individual color indices, we use parallelized color indices \citep[also see][]{lee19a,lee19c}. 
The RGB sequences in the individual color indices were parallelized using the following relation,
\begin{equation}
\parallel{\rm CI}(x) \equiv \frac{{\rm CI}(x) - {\rm CI}_{\rm red}}
{{\rm CI}_{\rm red}-{\rm CI}_{\rm blue}},\label{eq1}\label{eq:pl}
\end{equation}
where, CI$(x)$ is the color index of the individual stars and CI$_{\rm red}$, CI$_{\rm blue}$ are color indices for the fiducials of the red and the blue sequences of individual color indices \citep[see also][]{milone17}.

To perform a populational tagging of RGB stars in the magnitude range, \vvhbmag, we employed an expectation maximization (EM) algorithm for a 
multiple-component Gaussian mixture distribution model followed by the principal component analysis using the programming language $R$ \citep{rcore}. To make the best use of our measurements, we used \pcnjwl, \pchjwl, and \pnhjwl\ to calculate principal components as shown in Figure~\ref{fig:pca}.
We obtained the populational number ratio of \nrgb\ = 48:52 ($\pm$3) by employing the EM algorithm along the PC1, which is the exactly the same value that we obtained from the EM algorithm for the multiple-component Gaussian mixture distribution model along the \pcnjwl\ distribution \citep[see Table 5 of][]{lee19a}. Note that the eigenvector of the \pcnjwl\ is parallel to the PC1-axis that maximize the variance of the projected data, confirming our previous argument that \cnjwl\ is still important in populational tagging.

In order to circumvent potential confusion from population intruders and outliers in our analyses performed in \S\ref{s:tag} -- \S\ref{s:abund}, we used the well-behaved RGB stars located within 2$\sigma$ from the centers of each population on the PC1--PC2 plane as shown with dark gray color in Figure~\ref{fig:pca}.

As we noted in \citet{lee19a}, our populational number ratio of M3 RGB stars is significantly different from that by \citet{milone17}, who obtained an FG fraction of 0.305 $\pm$ 0.014 based on the \hst\ observations of the central part of the cluster. 
The discrepancy between our work and that by \citet{milone17} is solely due to a very strong radial gradient in the populational number ratio in M3, in the sense that the \cns\ population is more centrally concentrated and \citet{milone17} relied on the central part of the cluster, where the contribution of the \cns\ population is greater, while our study tends to be weighted more in the outer part of the cluster \citep[e.g., see Figure 8 of][]{lee19a}.

Since our photometry is incomplete in the central part of the cluster due to  limitations in spatial angular resolving power of the small aperture ground-based telescope, while that by \citet{milone17} is missing in the outer part of the cluster due to a small FOV of the \hst, we attempted to merge the data sets. Assuming our \cnw\ population is corresponding to the FG group by \citet{milone17}, we obtained the populational number ratio of \nrgb\ $\approx$ 38:62 ($\pm$2) from the merged data.

\begin{deluxetable*}{crcrrcrrcrrc}
\tablenum{2}
\tablecaption{
Pearson's correlation coefficients and $p$-values of the fit between our color indices} \label{tab:ci_corr}
\tablewidth{0pc}
\tablehead{\multicolumn{1}{c}{} & \multicolumn{5}{c}{\cnw} & 
\multicolumn{1}{c}{} & \multicolumn{5}{c}{\cns}\\
\cline{2-6}\cline{8-12}
\multicolumn{1}{c}{} & \multicolumn{2}{c}{\pcnjwl} & \multicolumn{1}{c}{} & \multicolumn{2}{c}{\pchjwl} &
\multicolumn{1}{c}{} & \multicolumn{2}{c}{\pcnjwl} & \multicolumn{1}{c}{} & \multicolumn{2}{c}{\pchjwl}\\
\cline{2-3}\cline{5-6}\cline{8-9}\cline{11-12}
\multicolumn{1}{c}{} & \multicolumn{1}{c}{$\rho$} & \multicolumn{1}{c}{$p$-value} & \multicolumn{1}{c}{} & \multicolumn{1}{c}{$\rho$} & \multicolumn{1}{c}{$p$-value} &
\multicolumn{1}{c}{} & \multicolumn{1}{c}{$\rho$} & \multicolumn{1}{c}{$p$-value} & \multicolumn{1}{c}{} & \multicolumn{1}{c}{$\rho$} & \multicolumn{1}{c}{$p$-value} 
}
\startdata
\pnhjwl & $-$0.060 & 0.206 &&    0.525 & 2.2$\times10^{-16}$ &&    0.153 & 7.9$\times10^{-4}$ && $-$0.008 & 0.849 \\
\pchjwl & $-$0.285 & 9.5$\times10^{-10}$ && \nodata & \nodata && $-$0.329 & 1.5$\times10^{-13}$ &&    \nodata & \nodata \\
\enddata 
\end{deluxetable*}

In panels (a), (b), and (c) of Figure~\ref{fig:cnchnh}, we show \pcnjwl\ versus \pchjwl, \pnhjwl\ versus \pchjwl, \pnhjwl\ versus \pcnjwl\ relations for M3 RGB stars in the magnitude range of \vvhbmag.
All of these relations exhibit discontinuities in the populational transition between the \cnw\ and \cns\ domains \citep[see also][]{smith13,lee17,lee18,lee19a,lee19c,lee20}. 
In no case does the photometric index comparison appear to connect the \cnw\ and \cns\ populations,
supported by the Pearson's correlation coefficients and $p$-values of the fit between our color indices as shown in Table~\ref{tab:ci_corr}.
Our results show that the chemical evolution from the \cnw\ population to the \cns\ population is not continuous in M3, already discovered in other GCs in our previous studies \citep{lee17,lee18,lee19c,lee20}.  
As we will discuss later, different index versus index trends between the \cnw\ and \cns\ populations shown in Figure~\ref{fig:cnchnh} are related to different [C/Fe] versus [N/Fe] relations, most likely due to different physical environments during the formation of each population.

Panels (d)-(f) of Figure~\ref{fig:cnchnh} show the variation of \pcnjwl, \pchjwl, and \pnhjwl\ indices with $V$ magnitude.
The sharp distinction between \cnw\ and \cns\ populations is easily seen in \pcnjwl\ at all giant branch luminosities.
However, \pchjwl\ and \pnhjwl\ do not show clear separation between the two groups.

\begin{deluxetable*}{crcrrccrcrrc}
\tablenum{3}
\tablecaption{
Pearson's correlation coefficients and $p$-values of the fit between our color indices and those of \citet{milone17}} \label{tab:hst}
\tablewidth{0pc}
\tablehead{\multicolumn{1}{c}{} & \multicolumn{5}{c}{\dtrio} & 
\multicolumn{1}{c}{} & \multicolumn{5}{c}{\dc}\\
\cline{2-6}\cline{8-12}
\multicolumn{1}{c}{} & \multicolumn{2}{c}{\cnw} & \multicolumn{1}{c}{} & \multicolumn{2}{c}{\cns} &
\multicolumn{1}{c}{} & \multicolumn{2}{c}{\cnw} & \multicolumn{1}{c}{} & \multicolumn{2}{c}{\cns}\\
\cline{2-3}\cline{5-6}\cline{8-9}\cline{11-12}
\multicolumn{1}{c}{} & \multicolumn{1}{c}{$\rho$} & \multicolumn{1}{c}{$p$-value} & \multicolumn{1}{c}{} & \multicolumn{1}{c}{$\rho$} & \multicolumn{1}{c}{$p$-value} &
\multicolumn{1}{c}{} & \multicolumn{1}{c}{$\rho$} & \multicolumn{1}{c}{$p$-value} & \multicolumn{1}{c}{} & \multicolumn{1}{c}{$\rho$} & \multicolumn{1}{c}{$p$-value} 
}
\startdata
\pnhjwl & $-$0.166 & 0.159 &&    0.010 & 0.308 &&    0.333 & 0.004 && $-$0.008 & 0.935 \\
\pchjwl & $-$0.346 & 0.002 && $-$0.614 & 7.7$\times10^{-13}$ &&    0.507 & 4.1$\times10^{-6}$ &&    0.422 & 3.9$\times10^{-6}$ \\
\pcnjwl &    0.296 & 0.011 &&    0.107 & 0.265 && $-$0.169 & 0.149 && $-$0.172 & 0.071 \\
\phkjwl & $-$0.191 & 0.102 &&    0.049 & 0.611 &&    0.574 & 9.1$\times10^{-8}$ &&    0.299 & 0.001 \\
\enddata 
\end{deluxetable*}

\section{Comparisons with \hst\ photometry}\label{s:hst}
\begin{figure*}
\epsscale{1}
\figurenum{6}
\plotone{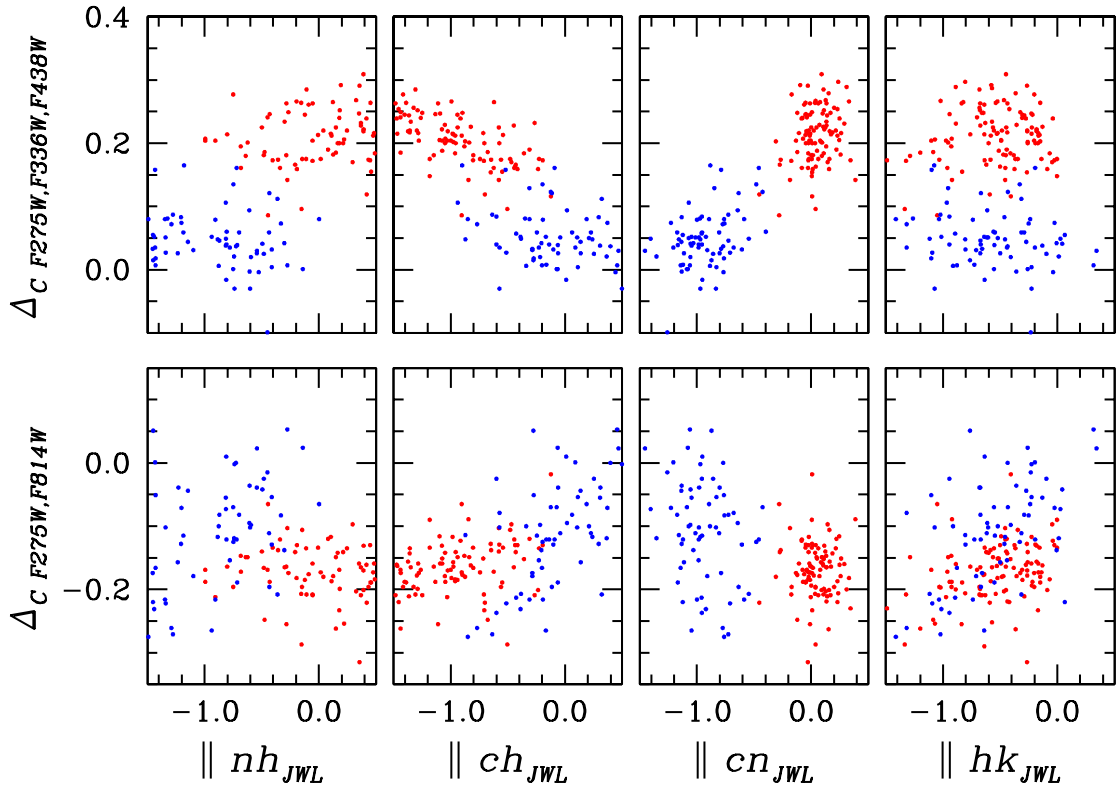}
\caption{
Comparisons of our color indices with those of \citet{milone17}. The blue and red dots denote the \cnw\ and \cns\ RGB stars in M3.
}\label{fig:hst}
\end{figure*}

We made comparisons of our color indices with those of the pseudo-color indices devised by \citet{milone15}, the \dtrio\ and \dc.
In Figure~\ref{fig:hst}, we show the results for stars with the radial distance of $r$ $\geq$ 1\arcmin\ to avoid the blending effect in our results. The figure indicates that the \cnw\ and \cns\ populations have different correlations with the \dtrio\ and \dc, which is not surprising because individual stellar populations on the so-called chromosome map of individual GCs of \citet{milone17} are not in linear relations. These different behaviors of individual populations have been discussed in previous paper of this series \citep{lee17,lee18,lee19c}. 
We calculated Pearson's correlation coefficients and $p$-values of the fit and we show our results in Table~\ref{tab:hst}. Our results are as follows;
\begin{itemize}
\item Our \pnhjwl\ index, i.e., the nitrogen abundance, does not appear to correlate with either \dtrio\ or \dc.
\item Our \pchjwl\ index, i.e., the carbon abundance, is anticorrelated with the \dtrio, while it is positively correlated with and \dc, with different degree of correlations between the \cnw\ and \cns\ populations.
\item Our \pcnjwl\ index is positively correlated with the \dtrio, which is consistent with the results from our previous works \citep{lee17,lee18}.
\item Finally, our \phkjwl\ (i.e., the metallicity index) of the \cnw\ population is correlated with the \dc\ in M3.
This suggests that the large extent of the \dc\ of the FG of stars in M3 pointed by \citet{lardo18} and \citet{tailo19} is likely due to the spread in metallicity, as we will discuss below.
\end{itemize}

\begin{figure}
\epsscale{1.}
\figurenum{7}
\plotone{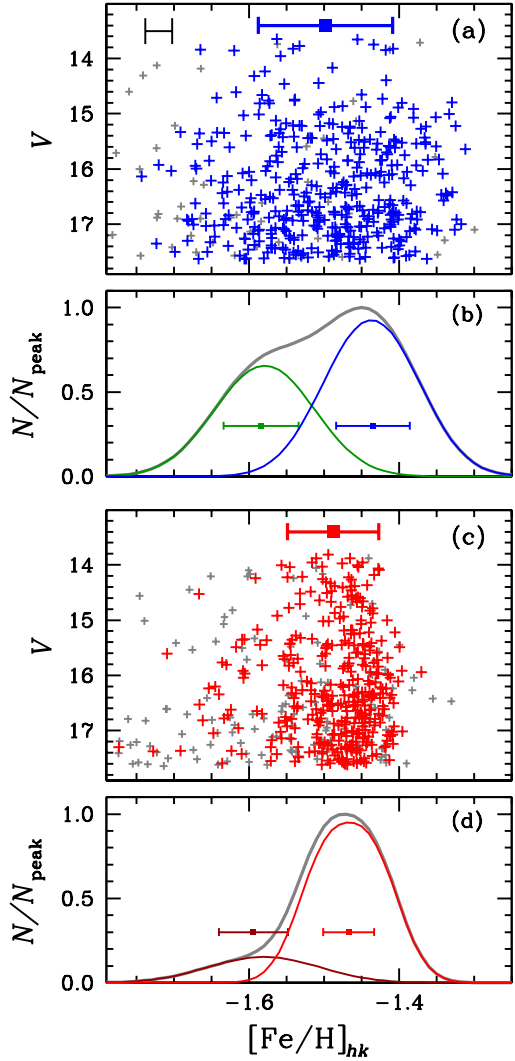}
\caption{
(a) Photometric metallicity from our \hkjwl, \fehhk, versus $V$ magnitude of the \cnw\ population. The blue plus signs denote the \cnw\ RGB stars with low measurement errors, $\sigma$(\hkjwl) $\leq$ 0.01 mag, while the gray plus signs indicate RGB stars with large measurement errors, $\sigma$(\hkjwl) $>$ 0.01 mag, those are not used in our analysis.
The blue filled square and the error bar indicate the mean \fehhk\ values and the standard deviation for the \cnw\ population shown with blue plus signs. The black error bar in the upper left corner denotes the $\pm1\sigma$ uncertainty in our \fehhk\ measurements.
(b) Generalized histograms returned from our EM estimator for each population. We also show the mean \fehhk\ values and standard deviations of each subpopulation. 
(c) Same as (a), but for the \cns\ population.
(d) Same as (b), but for the \cns\ population.}\label{fig:feh_hk}
\end{figure}

\begin{figure}
\epsscale{1.2}
\figurenum{8}
\plotone{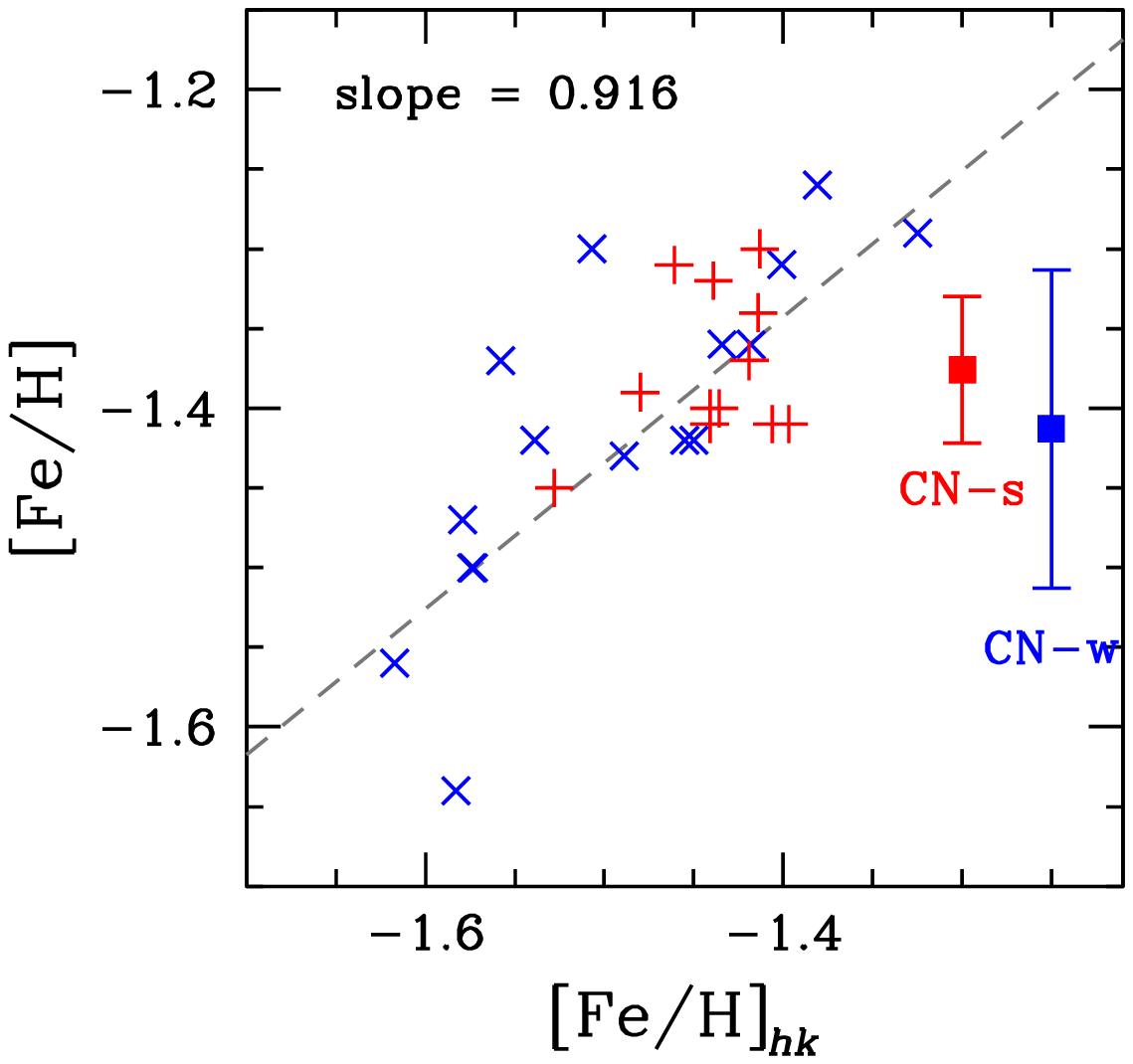}
\caption{
Comparison of our \fehhk\ with [Fe/H] of \citet{apogee}. The blue and red dots denote the \cnw\ and \cns\ RGB stars. The error bars indicate the mean [Fe/H] values and standard deviations of high-resolution spectroscopic measurements by \citet{apogee}.
}\label{fig:hk_FeHCaH}
\end{figure}

\section{Photometric Elemental Abundances}\label{s:abund}

\subsection{Synthetic Grids}\label{ss:grid}
We derive photometric [Fe/H], [C/Fe], and [N/Fe] from our photometric indices by comparing synthetic grids constructed using various input parameters. 

First, we obtained the Dartmouth model isochrones for [Fe/H] = $-$1.4, $-$1.5, and $-$1.6 with [$\alpha$/Fe] = +0.4, and the age of 12.5 Gyr \citep{dartmouth}. We interpolated the effective temperatures and surface gravities from $M_V$ = 3.5 to $-$2.5 mag with a magnitude step size of $\Delta M_V$ = 0.5 mag. Using these stellar parameters, we constructed series of synthetic spectra with varying carbon and nitrogen abundances with an abundance step size of 0.1 dex for carbon and 0.15 dex for nitrogen.

For our synthetic spectrum calculations we generated atomic/molecular line lists with the $linemake$ facility.\footnote{Available at \url{https://github.com/vmplacco/linemake}; linelists built by $linemake$ begin with those in the \cite{kurucz11} atomic line compendium and merge them with laboratory atomic line data from the University of Wisconsin and molecular line data from Old Dominion University. References to specific data sources are given in the $linemake$ web site.}
We computed the spectra with the 2011 version of the local thermodynamic equilibrium (LTE) line analysis code MOOG\footnote{Available at \url{http://www.as.utexas.edu/~chris/moog.html}.} 
\citep{moog,moogscat} that includes continuum scattering calculations.
For metal-poor stars, the negative hydrogen ion's bound-free (H$^-_{\rm bf}$) absorption dominates continuum opacities in the optical and infrared wavelength domains. 
However, in the blue and UV regions, Rayleigh scattering from neutral hydrogen atoms (RSNH) become the dominant source of continuum opacity, especially in cool RGB atmospheres \citep[e.g., see][]{suntzeff81,moogscat}.
Among many versions of MOOG, only that by \citet{moogscat} takes proper care of RSNH, which is important to calculate continuum opacities for our short-wavelength indices such as our $JWL34$ (the RSNH cross-section has a $\lambda^{-4}$ dependency).

Synthetic spectra can be calculated with the code SYNTHE \citep{kurucz05,castelli05}. It should be noted that the advantage of using MOOG is in its fast computations and its flexible treatment of input elemental isotopic ratios. 
For example, MOOG is several times faster than SYNTHE, which greatly reduced our total computation time for about 13,500 synthetic spectra with various input elemental abundances and stellar parameters. During our calculations, we compared continua returned from MOOG and SYNTHE for selective models and we found that both results are consistent. Therefore the results from our MOOG should be adequate for our purpose.

Finally, individual synthetic spectra were convolved with our filter transmission functions and then they were converted to our photometric system.

We emphasize that using our \hkjwl, \chjwl\ and \nhjwl\ makes  [Fe/H], [N/Fe] and [C/Fe] abundance estimates mathematically simple. At the depth of line formation, the nitrogen enhancement does not affect the CH molecule formation rate, while the carbon enhancement very weakly influences on the NH molecule formation rate through the enhanced formation of the CN molecule. Therefore, as we will discuss below, we can estimate [Fe/H], [C/Fe] and [N/Fe] as follows,
\begin{eqnarray}
{\rm [Fe/H]} &\approx& f_1(hk_{\rm JWL}, M_V),\label{eq:FeH} \\
{\rm [C/Fe]} &\approx& f_2(ch_{\rm JWL}, {\rm [Fe/H]}, M_V), \label{eq:CFe} \\
{\rm [N/Fe]} &\approx& f_3(nh_{\rm JWL}, {\rm [Fe/H]}, M_V),\label{eq:NFe}
\end{eqnarray}
where, [Fe/H] on the right hand side of Equations (\ref{eq:CFe}) and (\ref{eq:NFe}) are fed from Equations (\ref{eq:FeH}). The distance modulus and the interstellar reddening value for M3 are well known and, therefore, we can obtain accurate $M_V$ of individual RGB stars from our photometry. At given $M_V$, our synthetic model grids can be interpolated to match observed \hkjwl, \chjwl\ and \nhjwl\ to derive [Fe/H], [C/Fe] and [N/Fe].

Similarly, the [C/Fe] and [N/Fe] from the \cnjwl, \cfecn\ and \nfecn, can be given as
\begin{eqnarray}
{\rm [C/Fe]_{\rm CN}} &\approx& f_4(cn_{\rm JWL}, {\rm [Fe/H]}, {\rm [N/Fe]}, M_V), \label{eq:CFeCN}\\
{\rm [N/Fe]_{\rm CN}} &\approx& f_5(cn_{\rm JWL}, {\rm [Fe/H]}, {\rm [C/Fe]}, M_V), \label{eq:NFeCN}
\end{eqnarray}
where [N/Fe] and [C/Fe] on the right hand side of Equations (\ref{eq:CFeCN}) and (\ref{eq:NFeCN}) are fed from Equations (\ref{eq:NFe}) and (\ref{eq:CFe}), respectively.
At given $M_V$, [Fe/H], [N/Fe] and [C/Fe], our synthetic model grids were interpolated to match observed \cnjwl\ to derive \cfecn\ and \nfecn.

\begin{deluxetable*}{crrrr}
\tablenum{4}
\tablecaption{[Fe/H] of sub-populations and number ratios} \label{tab:subpop}
\tablewidth{0pc}
\tablehead{
\multicolumn{1}{c}{} & \multicolumn{2}{c}{[Fe/H]} & \multicolumn{1}{c}{} & \multicolumn{1}{c}{\nsub} \\
\cline{2-3}
\colhead{} &\colhead{SP1} & \colhead{SP2} & \colhead{} & \colhead{}}
\startdata
\cnw & $-$1.584 $\pm$ 0.050 $\pm$ 0.004 & $-$1.435 $\pm$ 0.049 $\pm$ 0.003 && 36:64 ($\pm$4)\\
\cns & $-$1.594 $\pm$ 0.046 $\pm$ 0.006  & $-$1.467 $\pm$ 0.034 $\pm$ 0.002 && 17:83 ($\pm$2)\\
\enddata 
\end{deluxetable*}

\subsection{Metallicity}\label{ss:feh}
We explore the metallicity spread of M3 using our \caii\ H \& K photometry. Calcium is synthesized in high-mass stars that die explosively, not in the kind of medium-mass stars whose AGB ejecta probably seeded the \cns\ stars of a GC's later population.
This makes our \hkjwl\ an excellent measure of metallicity with a fixed [Ca/Fe] abundance \citep{att91,lee09a,lee15}.
One of caveats for using the \caii\ H \& K lines is that these stellar lines arise from the \caii\ ground state, but so do absorption lines from gas in the interstellar medium. Thus interstellar contamination could affect the stellar absorption strengths.
In our previous study \citep[see Supplementary Information of][]{lee09a}, we demonstrated that the absorption of \caii\ H \& K lines by interstellar media having
a interstellar reddening of \ebv\ = 0.32 mag contributes only  $\Delta hk$ = 0.01 mag to our results. Since M3 does not show any perceptible differential reddening effect, the contribution of the interstellar \caii\ H \& K absorption can be completely neglected.

In Figure~\ref{fig:feh_hk}, we show plots of the [Fe/H]  derived from our \hkjwl\ index, \fehhk, versus $V$ magnitude for the \cnw\ 
and \cns\ populations. \hkjwl\ is an almost reddening-independent index, $E$(\hkjwl) = $-0.12\times E(B-V)$. 
Therefore the \fehhk\ dispersions shown in Figure~\ref{fig:feh_hk} cannot be due to differential reddening \citep{att91,lee09a}.
For this figure we selected only those RGB stars with very small measurement errors, $\sigma(hk) \leq$ 0.01 mag, in each population in order to ensure that their \hkjwl\ and subsequently \fehhk\ values are not affected by observational errors. 
Using Equation~\ref{eq:FeH}, we obtained \fehhk\ from our \hkjwl\ measurements. Metallicity gradients with $V$ are absent in both populations in the figure, which is natural since calcium is not synthesized or destroyed along the RGB phase of low-mass stars.
We obtained the mean photometric metallicities of $\langle$\fehhk$\rangle$ = $-$1.498 $\pm$ 0.089 $\pm$ 0.004 for 
the \cnw\ population and $-$1.488 $\pm$ 0.061 $\pm$ 0.003 for the \cns\ population, a nearly identical value (the errors denote  standard deviations and standard errors).

The most interesting and unexpected aspect of Figure~\ref{fig:feh_hk} is that our metallicity distributions for each population are well described by bimodal distributions.
In Table~\ref{tab:subpop}, we show the peak [Fe/H] values for individual sub-populations and the sub-populational number ratios returned from our EM estimators. The peak metallicities of sub-populations are in excellent agreement between the \cnw\ and \cns\ populations. However, the sub-populational number ratios are very different. The \cnw\ population has \nsub\ = 36:64 ($\pm$4), while the \cns\ population has 17:83 ($\pm$3), where the SP1 refers to the metal-poor component.

As we will argue later, the \cnw\ population shows a distinctive double RGBB distribution, which is related to the double metallicity distribution of the \cnw\ population.
On the other hand, we think that the conspicuous double RGBBs cannot be seen in the \cns\ population due to the low fraction of the metal-poor component, as we will discuss below.

Our results clearly show that the \fehhk\ (i.e., the \hkjwl) dispersion of the \cnw\ RGB stars is larger than that of the \cns\ population due to a low fraction of the metal-poor component of the \cns\ population, in accordance with the results from high-resolution spectroscopy that the metallicity spread of the \cnw\ population is larger than that of the \cns\ population \citep[e.g.,][]{sneden04,apogee}.

\citet{sneden04} studied 28 RGB stars in M3 using high-resolution visual spectroscopy. Unfortunately, since their sample stars are either too bright ($V$ $<$ \vvhb\ $-$ 2 mag) or too close to the cluster's center ($r <$ 1\arcmin), none of stars studied by \citet{sneden04} fall within our sample RGB selection criteria (\vvhbmag\ and $r \geq$ 1\arcmin). However, \citet{marino19} suggested that the metallicity spread for the \cnw\ population is larger. They obtained $\langle$[Fe/H]$\rangle$ = $-$1.63 $\pm$ 0.07 for FG, which is equivalent to the \cnw, and $-$1.57 $\pm$ 0.04 for SG, i.e., \cns, for M3 using the results by \citet{sneden04}.

More recently, \citet{apogee} performed high-resolution infrared spectroscopic study of M3. In Figure~\ref{fig:hk_FeHCaH}, we show comparisons of our \fehhk\ measurements with those of \citet{apogee}. As shown in the figure, our \fehhk\ measurements correlate well with [Fe/H] measurements by \citet{apogee}, confirming again that our \hkjwl\ is a good measure of metallicity.
We calculate Pearson's correlation coefficients and $p$-values of the fit, obtaining $\rho$ = 0.749 and $p$-value of 4.5$\times10^{-6}$.
Inspection of data by \citet{apogee} also suggests that the metallicity spreads of \cnw\ stars is larger than that of the \cns, and statistics of the two samples support this notion. 
For the \cnw\ group $\langle$[Fe/H]$\rangle$ = $-$1.41 $\pm$ 0.10 $\pm$ 0.03, while for the \cns\ group $\langle$[Fe/H]$\rangle$ = $-$1.38 $\pm$ 0.05 $\pm$ 0.01. We depict the [Fe/H] sample statistics with vertical lines drawn in Figure~\ref{fig:hk_FeHCaH} at arbitrary \fehhk\ positions for clarity. 
These result does not appear to conform to simple ideas of the sequential chemical evolution from the \cnw\ to the \cns\ populations, in which the later generation of stars, i.e., the \cns\ population in our study, would be naturally expected to have larger scatter in elemental abundances.

Finally, note in Figure~\ref{fig:hk_FeHCaH} that the number of the \cns\ stars studied by \citet{apogee} is smaller than that of the 
\cnw\ stars, \nrgb\ = 16:12. This may support our result that the \cns\ population is more centrally concentrated than the \cnw\ population is since the sample RGB stars studied by \citet{apogee} are located in the outer part of the cluster.

\begin{figure*}
\epsscale{1}
\figurenum{9}
\plotone{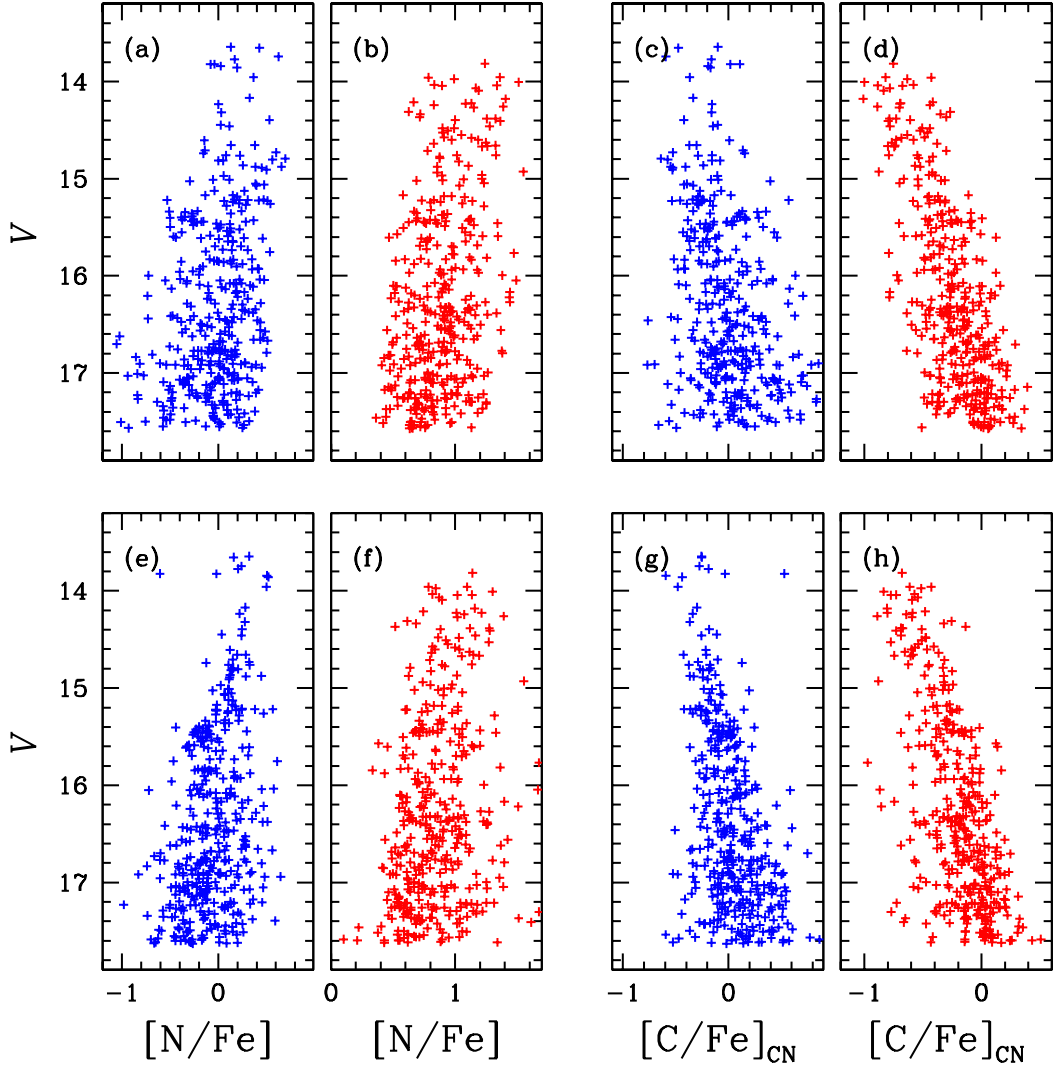}
\caption{
(a-b) [N/Fe] from \nhjwl. The blue and red color indicate the \cnw\ and \cns\ stars.
(c-d) [C/Fe] from \cnjwl\ with [N/Fe] from \nhjwl.
(e-f) Same as (a-b) but the metallicity effect is corrected.
(g-h) Same as (c-d) but the metallicity effect is corrected.
}\label{fig:nfe}
\end{figure*}

\begin{figure*}
\epsscale{1}
\figurenum{10}
\plotone{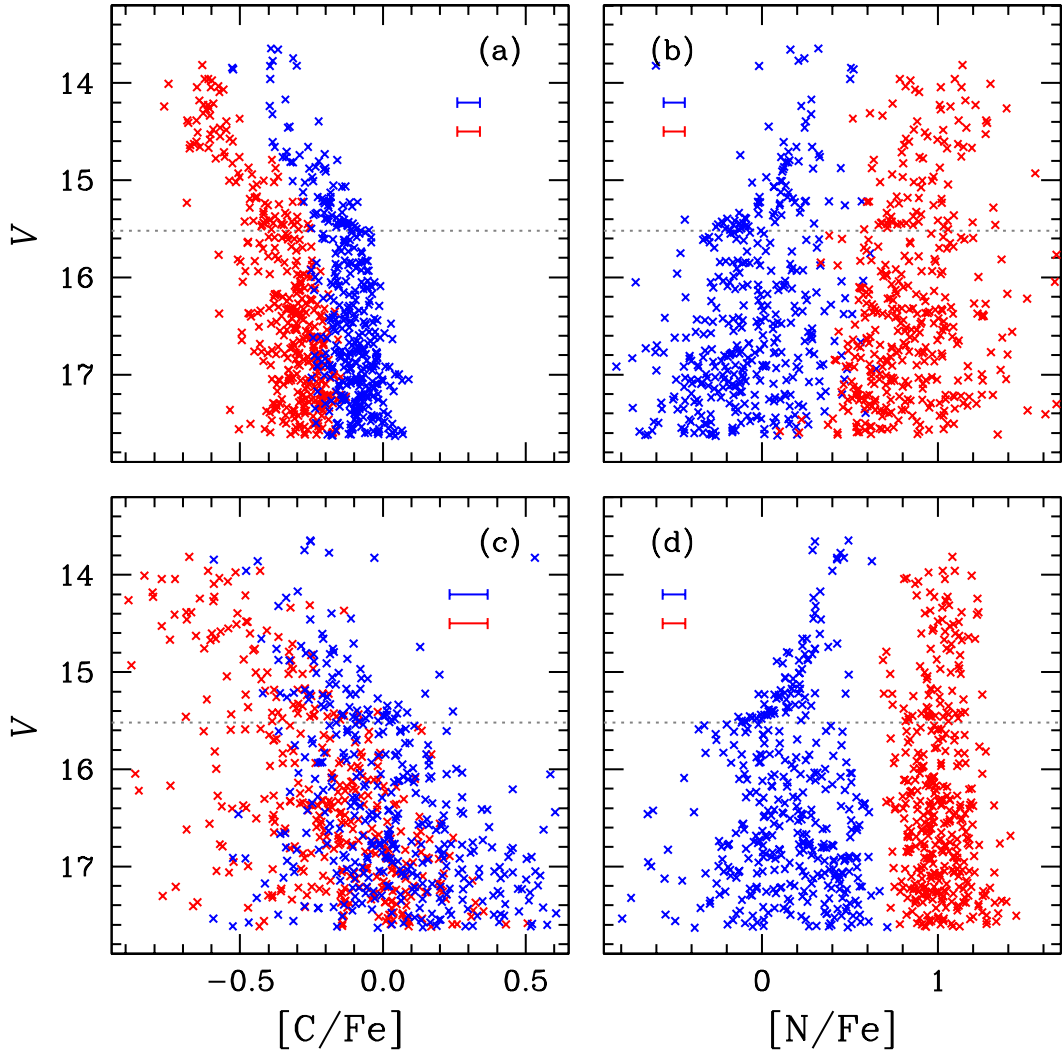}
\caption{
(a) [C/Fe] from \chjwl. The blue crosses are for the \cnw\ stars, while  the red crosses are for the \cns\ stars. The blue and red error bars denote measurement uncertainties ($\pm1\sigma$) for the \cnw\ and \cns\ stars, respectively.
(b) [N/Fe] from \nhjwl. 
(c) [C/Fe] from \cnjwl\ with [N/Fe] from panel (b) (i.e., [N/Fe] from \nhjwl).
(d) [N/Fe] from \cnjwl\ with [C/Fe] from panel (a) (i.e., [C/Fe] from \chjwl).
}\label{fig:CNabund}
\end{figure*}

\begin{figure}
\epsscale{1.2}
\figurenum{11}
\plotone{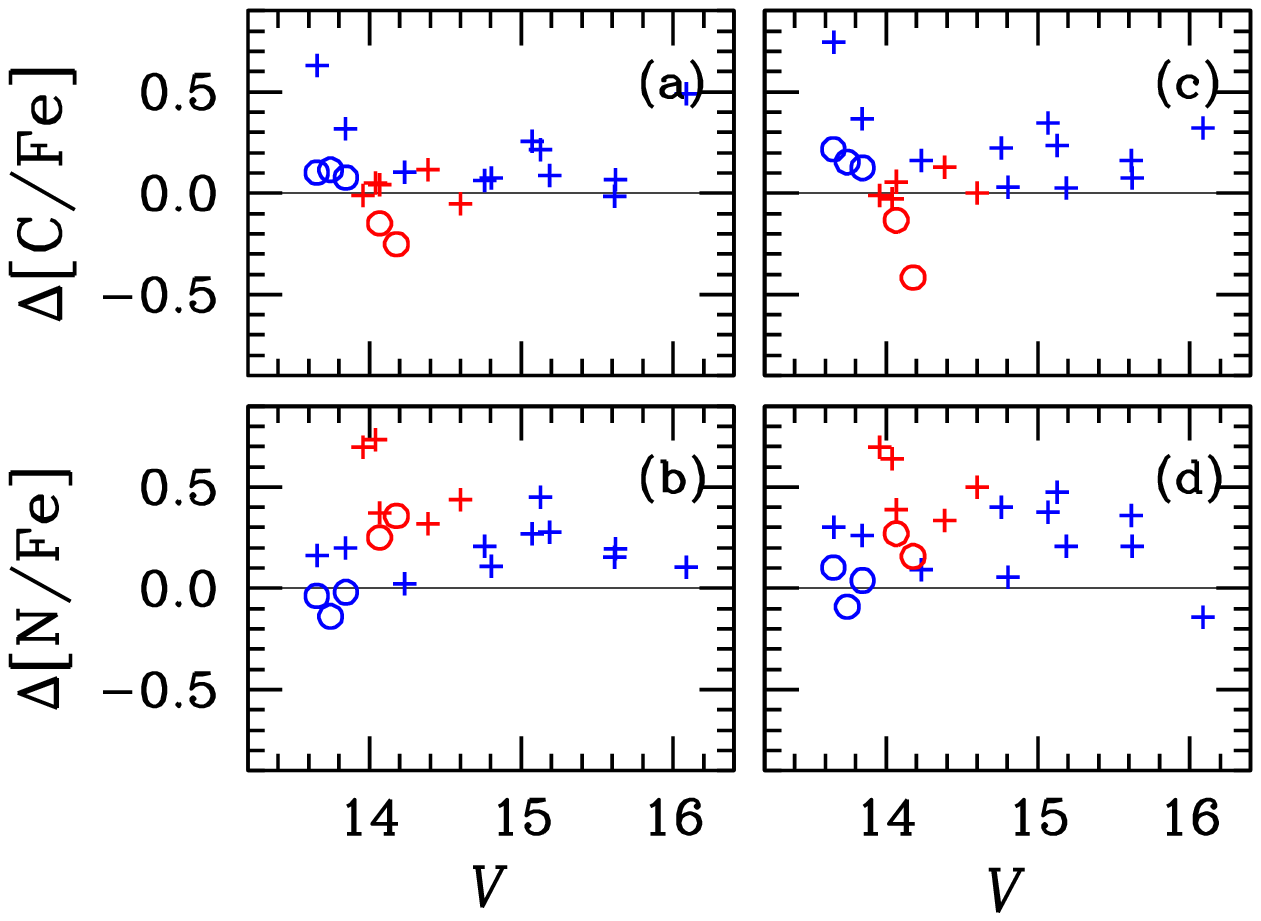}
\caption{
(a) Comparisons of [C/Fe] from \chjwl\ versus [C/Fe] by \citet[][open circles]{apogee}  and by \citet[][plus signs]{smith02}. The blue and red colors indicate the \cnw\ and \cns\ populations.
(b) Comparisons of [N/Fe] from \nhjwl\ versus [C/Fe] by \citet[][open circles]{apogee}  and by \citet[][plus signs]{smith02}.
(c) Same as (a) but for \cfecn.
(d) Same as (a) but for \nfecn.
}\label{fig:delCFeNFe}
\end{figure}

\begin{deluxetable*}{crrrrr}
\tablenum{5}
\tablecaption{Mean photometric [C/Fe] and [N/Fe] of M3 RGBs} \label{tab:CNabund}
\tablewidth{0pc}
\tablehead{
\multicolumn{1}{c}{} & \multicolumn{2}{c}{All} & \multicolumn{1}{c}{} & 
\multicolumn{2}{c}{fRGB($V$ $\geq$ \vbump)}\\
\cline{2-3}\cline{5-6}
\colhead{} &\colhead{[C/Fe]} & \colhead{[N/Fe]} & \colhead{} & \colhead{[C/Fe]} & \colhead{[N/Fe]}}
\startdata
\cnw & $-0.12\pm0.09\pm0.00$ & $-0.06\pm0.28\pm0.01$ && $-0.10\pm0.07\pm0.00$ & $0.10\pm0.28\pm0.01$\\
\cns & $-0.35\pm0.13\pm0.01$ & $0.85\pm0.26\pm0.02$ && $-0.30\pm0.08\pm0.00$ & $0.82\pm0.27\pm0.02$\\
\hline
\cnw\tablenotemark{1} & $0.02\pm0.32\pm0.02$ & $0.12\pm0.39\pm0.02$ && $0.06\pm0.33\pm0.02$ & $0.10\pm0.42\pm0.02$\\
\cns\tablenotemark{1} & $-0.21\pm0.26\pm0.01$ & $1.01\pm0.14\pm0.01$ && $-0.13\pm0.23\pm0.01$ & $1.01\pm0.14\pm0.01$\\
\enddata 
\tablenotetext{1}{\cfecn\ and \nfecn.}
\end{deluxetable*}

\begin{deluxetable*}{crrrrr}
\tablenum{6}
\tablecaption{Comparisons of photometric [C/Fe] and [N/Fe] to \citet{apogee} and \citet{smith02}} \label{tab:delCN}
\tablewidth{0pc}
\tablehead{
\multicolumn{1}{c}{} & \multicolumn{2}{c}{\citet{apogee}} & \multicolumn{1}{c}{} & 
\multicolumn{2}{c}{\citet{smith02}}\\
\cline{2-3}\cline{5-6}
\colhead{} &\colhead{[C/Fe]} & \colhead{[N/Fe]} & \colhead{} & \colhead{[C/Fe]} & \colhead{[N/Fe]}}
\startdata
\cnw & $ 0.10\pm0.02\pm0.01$ & $-0.07\pm0.05\pm0.03$ && $ 0.21\pm0.19\pm0.06$ & $0.20\pm0.11\pm0.03$\\
\cns & $-0.24\pm0.06\pm0.04$ & $ 0.13\pm0.03\pm0.02$ && $ 0.03\pm0.06\pm0.03$ & $0.51\pm0.17\pm0.07$\\
\cnw\tablenotemark{1} & $+0.20\pm0.05\pm0.04$ & $+0.30\pm0.05\pm0.04$ && $0.25\pm0.20\pm0.06$ & $0.24\pm0.17\pm0.05$\\
\cns\tablenotemark{1} & $-0.27\pm0.14\pm0.10$ & $+0.21\pm0.06\pm0.04$ && $0.03\pm0.06\pm0.03$ & $0.51\pm0.14\pm0.06$\\
\enddata 
\tablenotetext{1}{[C/Fe] and [N/Fe] from \cnjwl.}
\end{deluxetable*}

\begin{figure}
\epsscale{1.15}
\figurenum{12}
\plotone{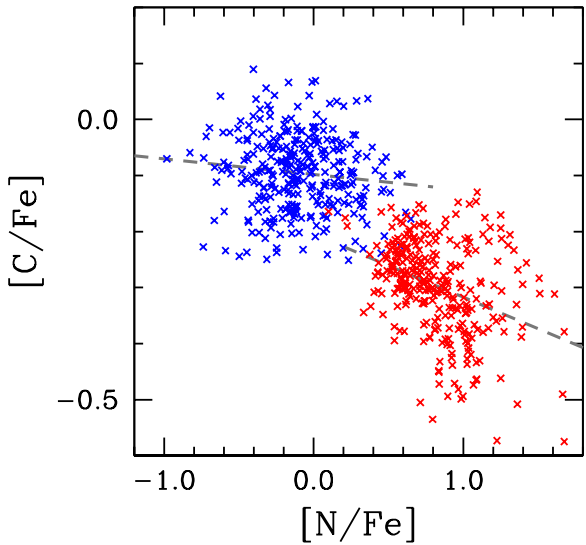}
\caption{
Plot of [C/Fe] versus [N/Fe] for M3 fRGB stars in the \cnw\ (blue) and the \cns\ (red) populations. The dashed lines show least-squares regression. The carbon abundance is weakly anticorrelated with (or remains flat against) the nitrogen abundance in the \cnw\ population, while the carbon and nitrogen abundances are more strongly anticorrelated in the \cns\ population.
}\label{fig:CFeNFe}
\end{figure}

\subsection{Carbon and Nitrogen}\label{ss:CN}
We derive photometric [C/Fe] and [N/Fe] abundances using our photometric measurements in two ways: (1) direct photometric [C/Fe] and [N/Fe] abundances from the \chjwl\ and \nhjwl\ indices using Equations~\ref{eq:CFe} and \ref{eq:NFe}, and (2) indirect photometric \cfecn\ and \nfecn\ abundances from \cnjwl\ with prior [C/Fe] and [N/Fe] measurements from the \chjwl\ and \nhjwl\ using Equations~\ref{eq:CFeCN} and \ref{eq:NFeCN}, respectively.

Carbon and nitrogen abundances derived from the CN band absorption strengths (as in our approach or as in low- to intermediate-resolution spectroscopy) depend on oxygen abundances due to the competition of CO molecules for carbon.
At a fixed carbon and nitrogen abundances, as oxygen abundance increases, the CO formation is slightly enhanced. Then the CN and CH formations are slightly suppressed due to decrease in available atomic carbon. On the other hand, the NH band strength (i.e., [N/Fe] from our \nhjwl) remains almost intact. Our calculations show that the NH populations at the optical depth of the line formation do not change with the variation of oxygen abundances.
The study of \citet{yong08} nicely illustrates the situation.
They obtained nitrogen abundances of a dozen RGB stars in NGC~6752 based on high-resolution spectroscopy of the NH band at \nhwave, and they argued that their nitrogen abundances are significantly different from those derived from the CN band by others. \citet{yong08} attributed the discrepancy in nitrogen abundances to inappropriate assignment of oxygen abundances during [N/Fe] derivation from the CN band by others. 

To compare our photometric carbon and nitrogen abundances with those of \citet{apogee}, we computed synthetic spectra assuming [O/Fe] = 0.50 for \cnw\ and 0.15 for \cns\ populations. 
These mean [O/Fe] values are consistent with those of three stars for the \cnw\ and two stars for the \cns populations by \citet{apogee}.
Note that their oxygen abundances are about 0.2--0.3 dex larger than those of \citet{sneden04}, who did not measure carbon and nitrogen abundances in M3. Using the results by \citet{sneden04}, we calculated $\langle$[O/Fe]$\rangle$ = 0.16 $\pm$ 0.06 (12 stars) for stars with [Na/Fe] $<$ 0, which are roughly corresponding to our \cnw\ population, and  $-$0.07 $\pm$ 0.12 for stars with [Na/Fe] $\geq$ 0 (i.e., $\approx$ \cns\ in our study). Our mean oxygen abundances are comparable to those of \citet{marino19}, who obtained $\langle$[O/Fe]$\rangle$ = 0.14 $\pm$ 0.04 (four stars) for the FG group and  0.00 $\pm$ 0.10 (eight stars) for the SG group using the results of \citet{sneden04}.

We are puzzled by this discrepancy since the oxygen abundance from the IR OH lines is supposed to be consistent with that from the [O I] line \citep[e.g., see][]{balachandran96}. The origin of this oxygen abundance discrepancy is beyond the scope of our current study. 
However, we caution that carbon and nitrogen abundances by \citet{apogee} could be incorrect since the oxygen abundances may affect the subsequent carbon and nitrogen abundance determinations through CO and CN bands.

Our \nfecn\ values are based on the same method as that employed in typical [N/Fe] derivations from CN bands used for low- to intermediate-resolution spectroscopy. But computations of \cfecn\ are less frequent in high-resolution spectroscopic studies.
Therefore we derived \cfecn\ values in order to check internal consistency between carbon abundances from the \chjwl\ and \cnjwl\ indices.

Our photometric [N/Fe] measurements from the \nhjwl\ index depend on metallicity since enhanced metallicity affects not only the line opacity of metallic lines, in particular in the shorter wavelength regime, but also the continuum opacity. As we will discuss in Appendix, $\Delta$[Fe/H] = 0.1 dex can result in $\Delta$[N/Fe] $\approx$ 0.1 -- 0.2 dex and the \cnw\ group is more vulnerable to such effect since it shows a substantial metallicity spread. We attempted to correct the metallicity effect on our photometric nitrogen abundance using our metallicity estimated from the \hkjwl\ already shown in Figure~\ref{fig:feh_hk}. 
In Figure~\ref{fig:nfe}, we show the [N/Fe] distributions with and without the metallicity correction, where a narrower RGB sequence in the [N/Fe] distribution of the \cnw\ can be seen  when a metallicity correction is applied. It is also true for the \cfecn\ of the \cnw, since the [N/Fe] is used to derive the \cfecn.
For the \cns\ population, the improvement is not as great as those can be seen in the \cnw\ population.

In Figure~\ref{fig:CNabund} and Table~\ref{tab:CNabund}, we show carbon and nitrogen distributions of M3 RGB stars. 
In the figure, the measurement uncertainties ($\pm1\sigma$) were estimated by the bootstrap method for 200,000 re-sampled data for each case.
For fRGB stars, we obtained $\langle$[C/Fe]$\rangle$ =  $-0.10\pm0.07\pm0.00$ and $\langle$[N/Fe]$\rangle$ = $0.10\pm0.28\pm0.01$ for the \cnw\ population, while $\langle$[C/Fe]$\rangle$ = $-0.30\pm0.08\pm0.00$ and $\langle$[N/Fe]$\rangle$ = $0.82\pm0.27\pm0.02$ for the \cns\ population. 
Both populations have nearly identical abundance dispersions for [C/Fe] and [N/Fe]. Also note that the carbon and nitrogen abundances from the CN band at \cnwave\ are higher than those from the CH and NH bands. If we attribute this discrepancy to uncertain oxygen abundances, our high \cfecn\ and \nfecn\ abundances may indicate that our assumed oxygen abundances from \citet{apogee} for each population ([O/Fe] = 0.5 dex for the \cnw\ and 0.15 dex for the \cns) are about 0.2 dex larger than they should be (see Figure~\ref{fig:ap:ofe}). If so, the resultant oxygen abundance ([O/Fe] = 0.30 dex for the \cnw\ and $-$0.05 dex for the \cns) are in good agreement with those of \citet{sneden04}.

We compare our photometric carbon and nitrogen abundances with those of \citet{apogee} and \citet{smith02}, showing our results in Figure~\ref{fig:delCFeNFe} and Table~\ref{tab:delCN}. In the table, we show standard deviations and standard errors for each case.
For the \cnw\ population, our carbon and nitrogen abundances are in good agreement with those of \citet{apogee}, while our [C/Fe] is slightly smaller and [N/Fe] is slightly larger than those of \citet{apogee} in the \cns. 
Our \cfecn\ and \nfecn\ are larger than [C/Fe] and [N/Fe] from the \chjwl\ and \nhjwl\ indices and, as consequences, they are larger than those of \citet{apogee} except for the \cfecn\ of the \cns\ population.
The differences between our results and those of \citet{smith02} are larger but this discrepancy is not a major issue.
As \citet{smith02} noted, their results were based on merged data in the literature and all of their [N/Fe] and most of their [C/Fe] were taken from \citet{suntzeff81}, whose input model atmospheres and spectral line information must have been very different than those used in our work.
We simply note that the absence of the gradient in the abundance difference against $V$ magnitude is encouraging. 

Finally, we show a plot of [N/Fe] versus [C/Fe] of the M3 fRGB stars in Figure~\ref{fig:CFeNFe}. We also calculated Pearson's correlation coefficients and $p$-values of the fit between the [C/Fe] versus [N/Fe] and we obtained $\rho$ = $-$0.117 and $p$-value of 0.041 for the \cnw\ and $\rho$ = $-$0.360 and $p$-value of 1.8$\times10^{-10}$ for the \cns\ populations.
The figure clearly shows the two separate relations between the carbon and nitrogen abundances. 
For the \cnw\ population, the carbon abundance is weakly anticorrelated with (or, perhaps, remains flat against) the nitrogen abundance, which may indicate that the \cnw\ fRGB stars formed out of gas experienced  no or little CN-cycle hydrogen burning, while the carbon and nitrogen abundances are more strongly anticorrelated in the \cns\ population, which is a natural consequence of the CN-cycle hydrogen burning.

\begin{figure}
\epsscale{1.15}
\figurenum{13}
\plotone{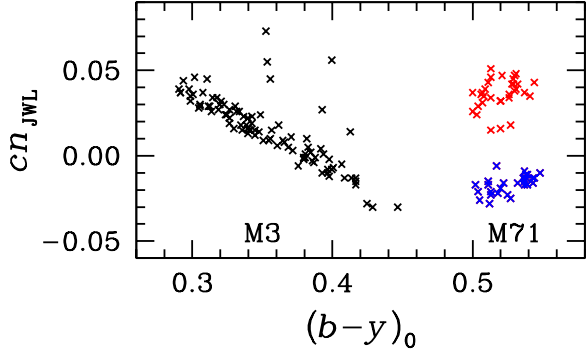}
\caption{
Plot of $(b-y)_0$ versus \cnjwl\ for RHB stars in M3 and the metal-rich GC M71. The \cnw\ and \cns\ M71 RHB stars are denoted with blue and red crosses, respectively. The six RHB stars separated far from the main body of the M3 RHB stars are thought to be the \cns\ RHB stars.
}\label{fig:rhb}
\end{figure}

\section{Populational Tagging for RHB}\label{s:rhb}

GC RHB stars are not warm enough to completely suppress the formation of diatomic molecules, such as NH, CN, and CH, and they can be used in populational tagging. For example, \citet{norris82} and \citet{smith89} studied CN bimodal distributions of RHB stars in metal-rich GCs, 47 Tuc and M71, respectively.

In Figure~\ref{fig:rhb}, we show a plot of $(b-y)_0$ versus \cnjwl\ for RHB stars in M3 and and M71. Note that the photometry for M71 is from our unpublished work. As shown in the figure, the RHB stars in metal-rich GC M71, [Fe/H] $\approx$ $-$0.8 dex, have a discrete bimodal \cnjwl\ distribution, which is consistent with the results by \citet[][see their Figures 2 and 3]{smith89}. Using our photometry, we obtained the RHB populational ratio of \nrgb\ = 46:54 ($\pm$11) for M71.

It is evident that the distribution of M3 RHB stars is different from that of M71. In particular the \cnjwl\ indices of M3 RHB stars increase with decreasing $(b-y)_0$ color. This is due to increasing contributions with effective temperature from H$_\zeta$ and H$_\eta$ at 3889.05\AA\ and 3835.38\AA, respectively, which reside in the $JWL38$ passband. The six RHB stars that deviate far from the main body of the M3 RHB stars are thought to be members of the \cns\ population.
If so, the M3 RHB stars are mainly composed of the \cnw\ population, 94 $\pm$ 3\%, which is consistent with the results from cumulative radial distributions as will be shown below. We also note that our results are in good agreement with the results of synthetic HB model simulations by \citet{tailo19} that the \cnw\ population is the major component for M3 RHB stars.

\begin{figure}
\epsscale{1.15}
\figurenum{14}
\plotone{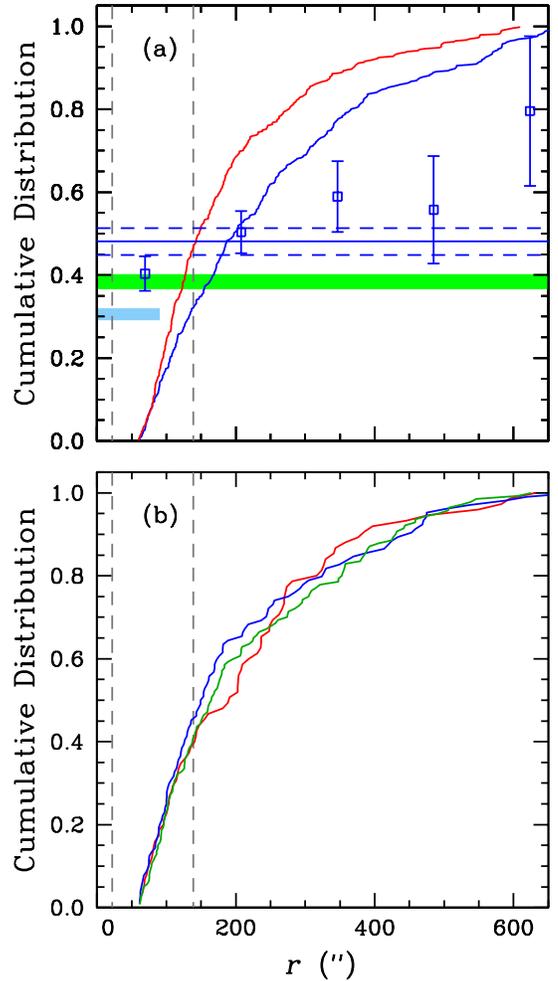}
\caption{
(a) CRDs of the \cnw\ (the blue solid line) and \cns\ (the red solid line) in the radial zone of $1\arcmin \leq r \leq 10\arcmin$. The blue open squares and error bars indicate mean \cnw\ fractions at given radial zones. The horizontal blue solid and dashed lines denote the mean fraction for the \cnw\ population with $\pm1\sigma$ and the vertical gray dashed lines denote the core and half-light radii of M3. The cyan shaded area indicates the fraction of the FG population with $\pm1\sigma$ and the FOV of \citet{milone17}, while the green shaded area indicates the fraction of the \cnw\ population estimated from merged data.
(b) CRDs of the RHB (the red solid line), RRL (the green solid line), and BHB (the blue solid line) stars in M3 with $r \geq$ 1\arcmin.
}\label{fig:cumrad}
\end{figure}

\begin{deluxetable}{lr}
\tablenum{7}
\tablecaption{$p$-values returned from the K-S Tests for CRDs with $r \geq$ 1\arcmin} \label{tab:KS}
\tablewidth{0pc}
\tablehead{\colhead{} &\colhead{$p$-value}}
\startdata
\cnw\ vs. \cns & 4.3$\times10^{-10}$ \\
& \\
\cnw\ vs. RHB & 0.367 \\
\cnw\ vs. RRL & 0.080 \\
\cnw\ vs. BHB & 0.008 \\
& \\
\cns\ vs. RHB & 0.029 \\
\cns\ vs. RRL & 0.181 \\
\cns\ vs. BHB & 0.724 \\
& \\
RHB vs. RRL & 0.646 \\
RHB vs. BHB & 0.249 \\
& \\
RRL vs. BHB & 0.893 \\
\enddata 
\end{deluxetable}

\section{Structural Differences between Multiple Stellar Populations}\label{s:struct}

\subsection{Cumulative Radial Distributions}\label{ss:crd}
The cumulative radial distributions (CRDs) of individual populations in GCs are frequently used to characterize populational properties and can also provide important information on their kinematics \citep[e.g.,][]{vesperini13}. 
In our previous study, we showed that the M3 \cns\ population is more centrally concentrated than the \cnw\ population and M3 has a strong radial gradient in the populational number ratio \citep{lee19a}. 

In Figure~\ref{fig:cumrad}, we show the CRDs of the \cnw\ and \cns\ RGB populations, RHB, RRL, and blue horizontal branch (BHB) stars in M3. 
We have not used stars in the central part of the cluster ($r$ $ \leq$ 1\arcmin), due to detection completeness issues.
We calculated the horizontal-branch ratio (HBR)\footnote{HBR = $(B-R)/(B+V+R)$, where $B, R, V$ denote the numbers of the BHB, RHB, and RRL, respectively.} from our photometry of HB stars with $r$ $ \geq$ 1\arcmin, and we obtained HBR = 0.08 $\pm$0.02, which is in excellent agreement with that of \citet[][the 2003 edition]{harris96}, 0.08.
We performed the Kolmogorov-Smirnov (K-S) tests between individual RGB and HB populations to see if they have the same CRDs and the $p$-value from our K-S tests are given in Table~\ref{tab:KS}. Our results are as follows;
\begin{itemize}
\item The \cnw\ and \cns\ RGB stars have completely different radial distributions.
\item The \cnw\ RGB and RHB are most likely drawn from the same parent distribution, suggesting that the bulk of the RHB stars are progeny of the \cnw\ RGB population. This is consistent with our result presented in \S\ref{s:rhb} that the major component of the M3 RHB is the \cnw\ population.  Similarly, \cns\ RGB and BHB are most likely drawn from the same parent distribution and the most of the BHB stars are most likely the progeny of the \cns\ RGB population. \item The \cnw\ RGB and BHB are not most likely drawn from the same parent distribution. Same is true for the \cns\ RGB and RHB stars.
\item The RRL stars are likely a mixed population; i.e., the progeny of both the \cnw\ and \cns\ RGB populations.
\item Unlike RGB populations, the RHB and BHB stars appear to share the similar CRD. It is possible that small numbers of the RHB and BHB stars may result in a slightly ambiguous $p$-value in our K-S test.
\end{itemize}

\begin{figure}
\epsscale{1.15}
\figurenum{15}
\plotone{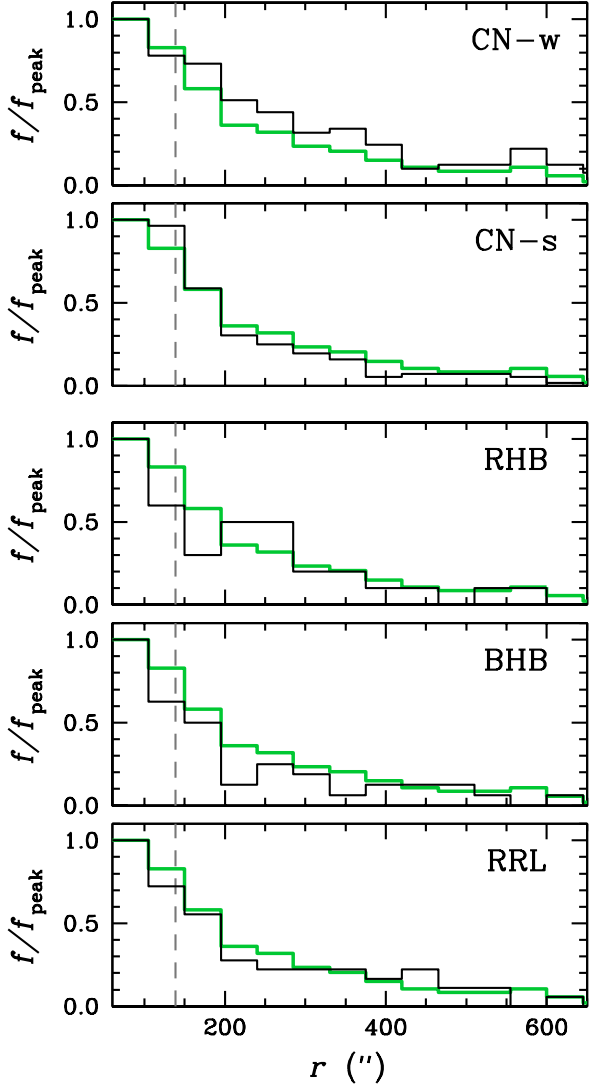}
\caption{
Radial distributions of RGB stars in M3. The green solid lines represent the histogram for the whole sample of stars. The vertical gray dashed lines denote the half-light radius of M3.
Note that  \cnw\ RGB and RHB populations appear to be slightly over-dense in the region between $r$ $\approx$ 200\arcsec\ and 300\arcsec.
}\label{fig:radial_no}
\end{figure}

The similarity in the structural properties between the \cnw\ RGB and RHB stars can also be found in Figure~\ref{fig:radial_no}. 
In the figure, we show histograms of individual populations against the radial distance. 
In a radial zone from $r$ $\approx$ 200\arcsec\ and 300\arcsec, both the \cnw\ RGB and RHB stars are slightly over populated than the average distribution of stars, a strong indication that the complete homogenization had not been achieved in the \cnw\ population.

\begin{figure}
\epsscale{1.15}
\figurenum{16}
\plotone{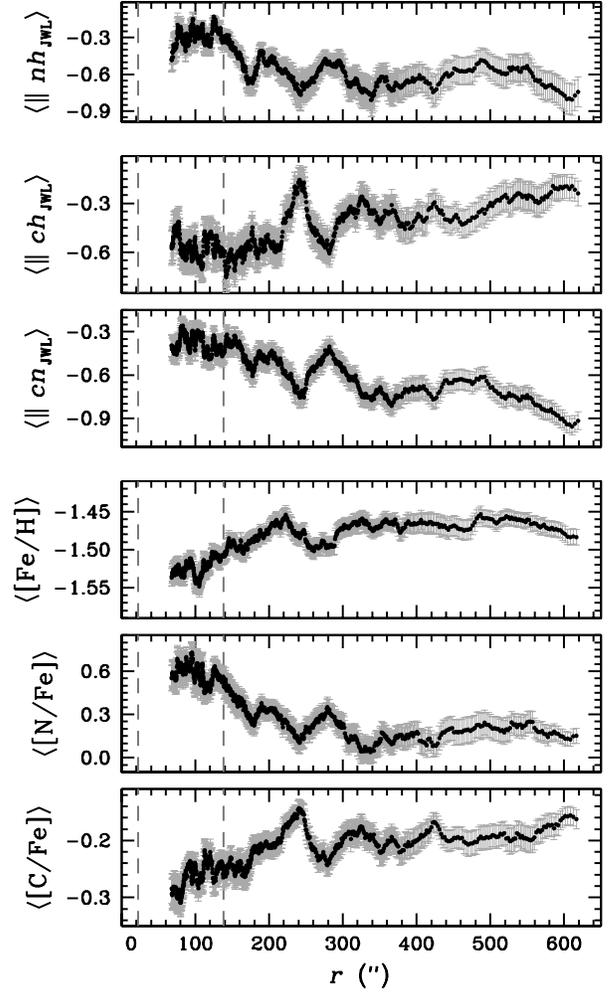}
\caption{
Moving averages of the adjacent 50 points for each color index of the RGB stars plotted as functions of radial distance from the M3 center. The vertical thin gray error bars denote the standard error of the mean and the vertical gray dashed lines denote the core and the half-light radii of M3.
}\label{fig:radial_ci}
\end{figure}

In Figure~\ref{fig:radial_ci}, we show the moving averages of the RGB stars for the adjacent 50 points of the \pnhjwl, \pchjwl, and \pcnjwl\ indices and \fehhk, [N/Fe] and [C/Fe] against radial distance. 
The figure shows that M3 exhibits a metallicity gradient against the radial gradient, which is an unexpected result. 
In Section~\ref{s:merger}, we will discuss that M3 appears to consist of the two GCs with slightly different metallicities, namely the C1 and C2. As we will show later, this metallicity gradient in M3 is consistent with the idea that the metal-poor C1 is more centrally concentrated than the metal-rich C2.
On the other hand, due to the domination of the \cns\ population in the central part of M3 as already shown in Figure~\ref{fig:cumrad}, the large values of the \pnhjwl\ and \pcnjwl\ index and the small value of the \pchjwl\ index and, subsequently, the large value of the [N/Fe] and the small value of the [C/Fe] can be found in the inner part of the clusters, reminiscence of the strong radial CN variation in 47~Tuc \citep{chun79}. 
From the radial distance of about 200\arcsec\ to 300\arcsec, bumps in the all color indices and the \fehhk, [N/Fe] and [C/Fe] are noticeable due to the existence of the over-population of the \cnw\ RGB stars in this radial zone as we already showed in Figure~\ref{fig:radial_no}.

\subsection{Spatial Distributions}\label{ss:spatial}
The GC morphology can be affected by both internal processes, such as relaxation and cluster rotation, and external processes, such as tidal fields.
For example, the ellipticity of GCs can be affected by the dynamical relaxation that leads the stars with excess angular momentum to evaporate \citep[e.g.,][]{shapiro76}.
On the other hand, \citet{bianchini13} argued that the well-relaxed GC 47~Tuc can be explained very well with its internal rotation.

Here, we investigate the morphology of each population. If both populations were dynamically evolved together, the dynamical relaxation and the tidal effect should be similar for both populations. 
Therefore, it can be thought that morphological difference between the two populations is due to other effects, such as internal rotation, as we will discuss below.
The N-body simulations by \citet{vesperini13}, for example, showed that the complete mixing of GCs can be achieved in about at least 20 half-mass relaxation time. Since the half-mass relation time for M3 is about 2 Gyr \citep[][2003 version]{harris96}, the complete mixing has not been achieved in M3 within the Hubble time and the two stellar populations in M3 may keep their initial kinematical properties.

\begin{figure}
\epsscale{1.15}
\figurenum{17}
\plotone{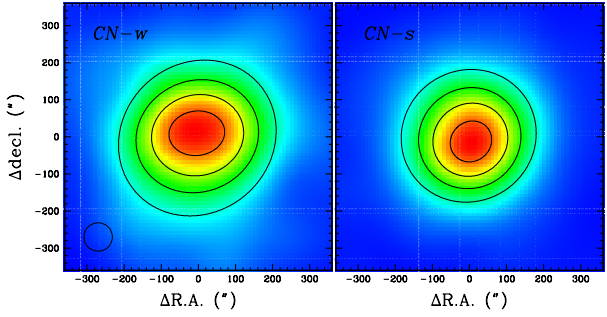}
\caption{
Smoothed density maps of the \cnw\ (left) and \cns\ (right) RGB populations. The FWHM of our Gaussian kernel are shown with a circle and isodensity contour lines for 90\%, 70\%, 50\%, and 30\% of the peak values of each population are also shown.  Note that the absolute peak values are different for each population.
}\label{fig:spatial}
\end{figure}

\begin{figure}
\epsscale{1.15}
\figurenum{18}
\plotone{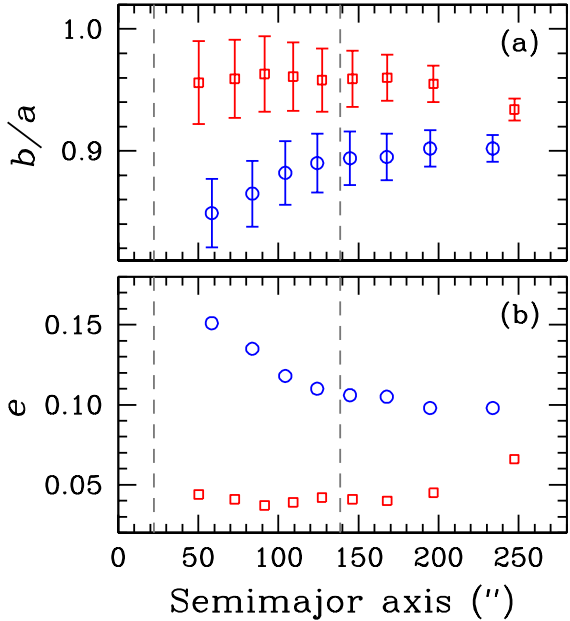}
\caption{
The run of the axial ratio $b/a$, and ellipticity $e$, of the \cnw\ (blue) and \cns\ (red) populations in M3 against the semimajor axis $a$. The vertical gray dashed lines denote the core and half-light radii of M3. The \cnw\ population is more elongated.
}\label{fig:ellip}
\end{figure}

In order to investigate the spatial distributions of the \cnw\ and \cns\ RGB populations, we constructed smoothed density maps following the same method described in our previous work \citep[][and see references therein]{lee15}. 
We present our results in Figure~\ref{fig:spatial}, showing that the \cnw\ population has a more elongated spatial distribution. 
The evidence of the spatially more elongated structure of the \cnw\ population can be clearly seen in Figure~\ref{fig:ellip}, where we show radial distributions of the axial ratio, $b/a$, and the ellipticity, $e=(1-b/a)$, of the \cnw\ and \cns\ populations. 
As shown in the figure, the ellipticity of the \cnw\ population is significantly larger than the \cns\ population up to $r$ $\approx$ 1.5$r_h$. At large radial distance, the axial ratios of the \cnw\ and \cns\ appear to converge on $b/a$ $\approx$ 0.90 -- 0.94, consistent with that of \citet{chen10}, who obtained the axial ratio of $b/a = 0.94 \pm 0.01$ at $a \approx$ 11\arcmin\ for M3 using the 2MASS points sources. 

\subsection{Internal Rotations}\label{ss:rot}
Some GCs, such as 47~Tuc, M15, and $\omega$ Cen, show substantial internal rotations, reaching the maximum line-of-sight velocities of $\approx$ 3 -- 6 \kms. 
Previous studies of M3's rotation using radial velocities showed that M3 has a weak rotational component, in the order of 1.0 -- 1.5 \kms\ \citep{fabricious14,ferraro18}.
Note that the internal rotation of GCs is not an invariant parameter but it can change with time. For example, numerical simulations by \citet{wang16} showed that the internal rotation of GCs can gradually decrease with time due to the loss of angular momentum via the two-body relaxation and mass loss, although the timescale for such processes appears to be somewhat uncertain.

In our previous studies we reported that the structural and kinematical properties between MPs in GCs are decoupled \citep[e.g., see][]{lee15,lee17,lee18,lee20}.
In the same spirit we have now investigated the tangential rotation of the \cnw\ and \cns\ populations using the proper motion study of the \gaia\ DR2 \citep{gaiadr2}.

\begin{figure}
\epsscale{1.2}
\figurenum{19}
\plotone{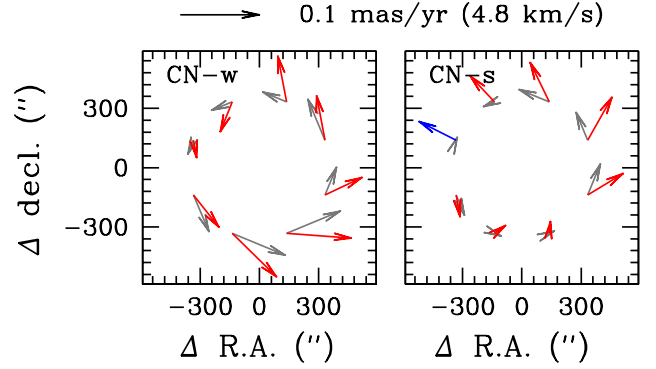}
\caption{
Mean proper motion vectors of eight slices in the radial zone of 1\arcmin\ $\leq$ $r$ $\leq$ 10\arcmin. The red arrows denote a counterclockwise rotation (E $\rightarrow$ N $\rightarrow$ W $\rightarrow$ S $\rightarrow$ E), while the blue arrow denote a clockwise rotation at a given position vector. The gray arrows represent the rotating component assuming a circular rotation. An arrow shown in above indicates a proper motion of 0.1 mas/yr or 4.8 \kms\ assuming 10.2 kpc for the distance to M3.
}\label{fig:rot}
\end{figure}

We divided the sphere into eight different slices in a single radial zone of 1\arcmin\ $\leq$ $r$ $\leq$ 10\arcmin, and we calculated the mean proper motion vectors in each slice. 
In Figure~\ref{fig:rot}, we show the tangential rotation of each population. We note that both populations share the same direction in their rotation. On the celestial sphere, they appear to rotate in a counterclockwise sense (E $\rightarrow$ N $\rightarrow$ W $\rightarrow$ S $\rightarrow$ E) with the \cnw\ population having a more well-defined tangential rotation as shown with gray arrows.
Assuming a circular projected rotation for both populations, we calculated mean rotational components of 0.068 $\pm$ 0.031 mas and 0.033 $\pm$ 0.029 mas for the \cnw\ and \cns\ populations, respectively, which are corresponding to 3.28 $\pm$ 1.51 \kms\ and 1.60 $\pm$ 1.41 \kms, respectively, assuming the distance to M3 of 10.2 kpc \citep{harris96}.
Our results show that the \cnw\ population exhibits a larger, statistically significant, net tangential rotation than the \cns\ population, suggesting that its large ellipticity in the M3 \cnw\ population is likely induced by its fast internal rotation, as can be seen in M5, M22, and NGC 6752 in our previous studies \citep[][and references therein]{lee15,lee17,lee18}.

\begin{figure*}
\epsscale{1}
\figurenum{20}
\plotone{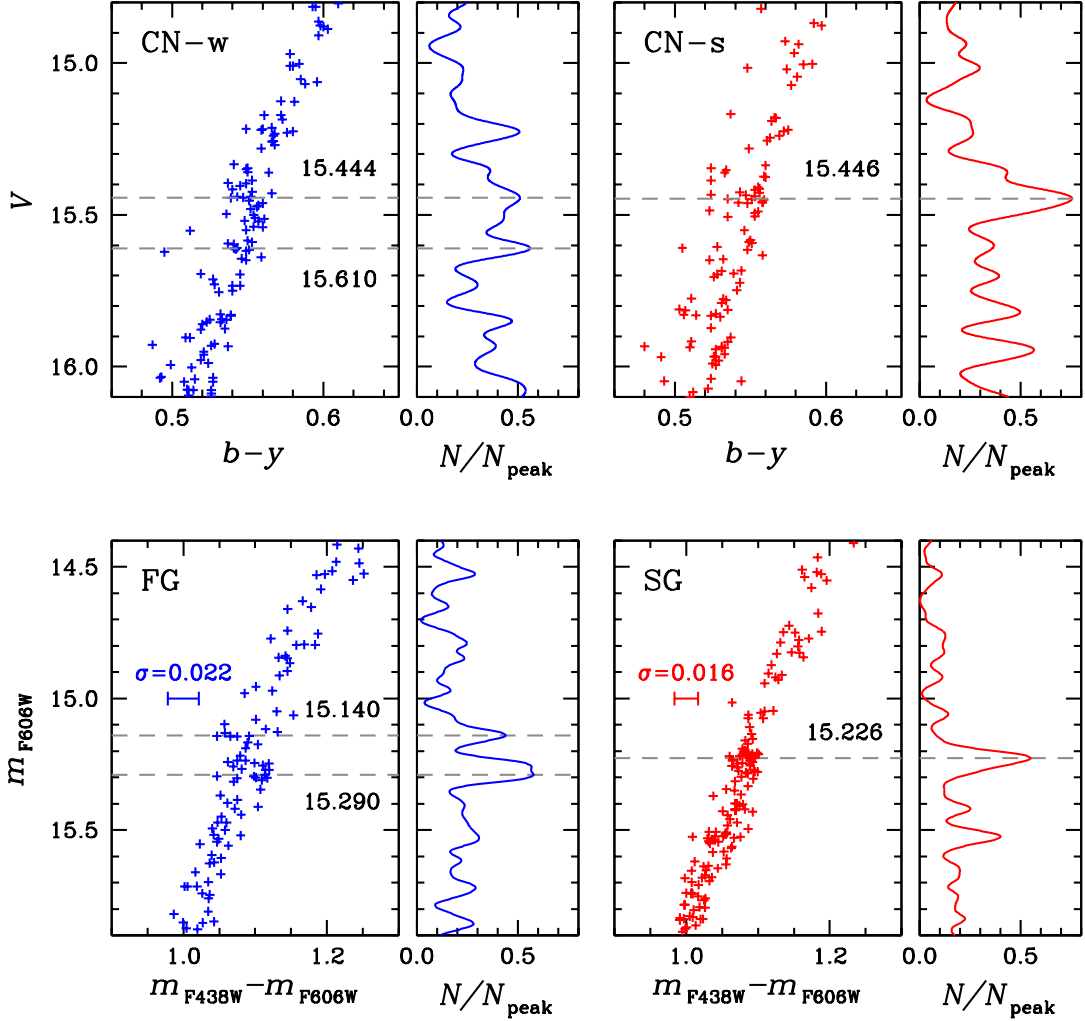}
\caption{
(Top panels) CMDs and differential LFs of M3 from our observations. The blue and red colors denote the \cnw\ and \cns. The gray dashed lines represent the $V$ magnitude of the RGBBs. (Bottom panels) CMDs and differential LFs of M3 using the \hst\ photometry by \citet{uvlegacy}.
The $m_{\rm F606W}$ values for RGBB are shown with gray dashed lines. The RGB widths around the RGBB region are also shown with error bars.
}\label{fig:rgbb}
\end{figure*}

\section{The Double Red Giant Branch Bumps of the M3 \cnw}\label{s:rgbb}

In the course of low mass star evolution, RGB stars experience slower evolution and temporary drop in luminosity when the very thin H-burning shell crosses the discontinuity in the chemical composition and lowered mean molecular weight left by the deepest penetration of the convective envelope during the ascent of the RGB, the so-called RGBB \citep[e.g., see][]{cassisi13}. 
It is well known that the RGBB luminosity increases with helium abundance and decreases with metallicity at a given age \citep[e.g., see][]{bjork06,valcarce12}.

The helium abundance of GC stars cannot be measured directly due to the lack of helium absorption lines.
Instead, the RGBB magnitude can be a powerful tool to probe the helium content of a given stellar group at a fixed age and metallicity. 
For example, \dy\ $\approx$ 0.02 can be translated into $\Delta$[He/H] $\approx$ 0.03 dex, which is almost impossible to detect in high resolution spectroscopy. On the other hand, this small helium difference can affect the RGBB magnitude by $\Delta V$ $\approx$ 0.05 mag, which can be easily detected, unless there exists no degeneracy with metallicity \citep[e.g., see][]{lee15,lee17,lee18,lee19c,lagioia18,milone18}.

As mentioned above, in M3 the extent of the \dc\ distribution of the \cnw\ population (i.e., the FG of stars in the frame of the sequential formation of the MPs in GCs) is unusually large \citep{milone17,lardo18}.
Based on the dependence of synthetic color indices with varying He abundances but a fixed metallicity, \citet{lardo18} claimed that the large extent of the \dc\ distribution of the M3 \cnw\ RGB stars is due to the helium enhancement as large as $\Delta$Y $\approx$ 0.024 and a small range of nitrogen abundance dispersion within the \cnw\ population. However, our results show that the \cnw\ and \cns\ have identical nitrogen abundance dispersion as shown in Table~\ref{tab:CNabund}, $\sigma$[N/Fe] $\approx$ 0.28 dex.

More recently, \citet{tailo19} undertook synthetic HB model simulations, finding that a model without helium enhancement in the FG of stars can best match with the observed M3 HB type, RRL period distribution, and the MS distribution. 
However, they noted that they were not able to address the color spread of the M3 FG star satisfactorily. \citet{tailo19} also did not consider the metallicity spread among the M3 FG stars.

The large helium spread of $\Delta$Y $\approx$ 0.024 in the FG of stars without any perceptible CNO abundance spreads by \citet{lardo18} is difficult to reconcile with the current understanding of the GC formation scenario. 
As \citet{lardo18} argued, their results imply that $p-p$ chain hydrogen burning induced helium enhancement within the FG of stars. 
If so, the FG of a such GC requires a very long time scale and it does not conform to the currently believed sequential two-step formation scenario with a relatively short time scale, $\lesssim$ 100 Myr, or the small age dispersion in M3 seen from the \hst\ observations, $\pm$0.5 Gyr \citep{dotter10}.

Here, we investigate the helium content and metallicity of  M3 \cnw\ population using the RGBB. 
From a stellar evolution perspective the RGBB is less complicated than the HB and AGB. 
For example, the HB morphology of GCs depends not only on the helium abundance but also on the mass-loss during the RGB phase. 
At the same time, typical GC HB stars are too faint to perform populational tagging from high-resolution spectroscopy.
Tagging HB stars from low-resolution spectroscopy or from narrow band photometry, such as ours, would also be difficult for hot BHB stars, where the most informative diatomic molecular bands are absent.
In addition to statistical fluctuations due to small number statistics of AGB stars, the unsolvable problem with populational number ratio study of AGB stars is the missing populations that eventually evolve in to the \agbm\ \citep[e.g., see Figures 20 and 24 of][]{lee17}.
On the other hand, the populational tagging for the entire RGBB populations can be readily done with our new color indices.

\begin{figure}
\epsscale{1.2}
\figurenum{21}
\plotone{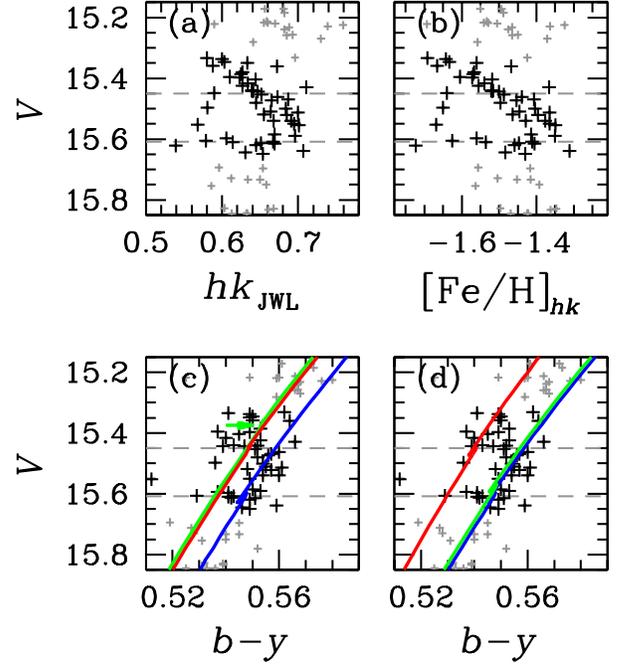}
\caption{
(a) \hkjwl\ versus $V$ CMD around the RGBB region of the \cnw\ population. The gray dashed lines indicate the $V$ magnitudes of the double RGBBs and black plus signs denote the stars near the double RGBBs.
(b) Same as (a), but for \fehhk.
(c) $(b-y)$ versus $V$ CMD. The blue solid line represents the model isochrone for Y = 0.25 and [Fe/H] = $-$1.45 dex, while the red solid line represents that of [Fe/H = $-$1.50 dex and Y = 0.29. The green solid line and the arrow indicate the isochrone and the RGBB $V$ magnitude for [Fe/H = $-$1.45 dex and Y = 0.33.
(d) The blue and red solid lines represent the model isochrones for Y = 0.25 and [Fe/H] = $-$1.45 and $-$1.60 dex, respectively, while the green solid line represents that for Y = 0.26 and [Fe/H] = $-$1.45 dex.
}\label{fig:bumpiso}
\end{figure}

\subsection{Double RGBBs of the M3 \cnw}\label{ss:double_rgbb}
As we mentioned above, the extended and tilted RGBB os the \cnw\ population is one of the interesting features of M3. 
In our current work, we investigate the the differential luminosity functions (LFs) of each population to understand the underlying stellar populations in M3.
In Figure~\ref{fig:rgbb}, we show CMDs and LFs of both populations, finding \vbump\ = 15.610 $\pm$ 0.025 mag  and 15.444 $\pm$ 0.025 mag for the \cnw\ population and 15.446 $\pm$ 0.025 mag for the \cns\ population. Using all RGB stars from both populations, we obtained 15.445 $\pm$ 0.02 mag and our RGBB $V$ magnitude is in excellent agreement with that of \citet{ferraro97}, who reported \vbump\ = 15.45 $\pm$ 0.05 mag for M3.
The double RGBBs of the M3 \cnw\ population from our ground-based observations can also be supported by the \hst\ photometry by \citet{uvlegacy}. In Figure~\ref{fig:rgbb}, we also show CMDs and differential LFs of the FG and SG groups. We note again that the FG and SG groups are corresponding to our \cnw\ and \cns\ populations, respectively. The figure clearly shows that the FG group has discrete double RGBBs, consistent with our results for the \cnw\ population. Interestingly, the magnitude difference between the two RGBBs of the \hst\ photometry is 0.150 $\pm$ 0.020 mag, in excellent agreement with our ground based observation, 0.166 $\pm$ 0.035 mag. 
We derived fourth order polynomial fits to measure the RGB width of the FG and SG groups. We calculated scatters around the fitted lines and we obtained $\sigma$($m_{\rm F438W} - m_{\rm F606W}$) = 0.022 mag for the FG and 0.016 mag for the SG. The broader RGB width of the FG group is  consistent with our previous result shown in Figure~\ref{fig:feh_hk}.

We note that the RGBB magnitude of the SG group from the \hst\ photometry does not agree with our results. The \cns\ RGBB magnitude is almost identical to that of the bright \cnw\ RGBB. However, the RGBB magnitude of the SG group from the \hst\ observations is 0.086 mag fainter than the bright RGBB of the FG group.
The cause of this discrepancy is unclear. 
(1) It could originate from the different behavior in different photometric systems with significantly different bandwidths (i.e., the \str\ $y$ versus the \hst\ F606W) for certain chemical abundances. 
But our calculations show that the difference in the filter passbands between the \str\ $y$  and the \hst\ F606W can only explain the magnitude difference of 0.005 mag. 
(2) Perhaps there might exist a radial gradient of the \cns\ RGBB in the sense that the \cns\ RGBB gets fainter with the decreasing radial distance, probably due to increasing metallicity of the main body of the \cns. We examined our photometry within the radial zone of 1\arcmin\ $\leq$ $r$ $\leq$ 10\arcmin, but we did not find any hint of the radial gradient of the \cns\ RGBB magnitude. It is interesting to note that the double RGBBs cannot be seen in the \cns\ population, which we will discuss later. Future studies of this issue would be welcome.

At face value, if we assume that only helium abundance affects the RGBB magnitude, the difference in the magnitude between the faint \cnw\ RGBB and \cns\ RGBB, $\Delta V$ = 0.164 $\pm$ 0.035 mag in our observations can be translated into a helium spread of \dy\ = 0.066 $\pm$ 0.014, in the sense that the \cns\ is significantly enhanced in helium abundance, which does not seem very plausible for the M3 HB morphology.

This is an apparently straightforward conclusion, but the interpretation of the detailed sub-structures of the RGBB may not be so simple. 
The double RGBBs in the \cnw, with the magnitude difference between the two peaks of \dvbump\ $\approx$ 0.16 mag, can arise from at least three different physical sources: The differential reddening effect, the helium and metallicity spreads.

\subsection{Differential Reddening}\label{ss:rgbb_red}
Differential reddening across the M3 field could lead to an apparent double RGBBs, although no such effect has been reported so far. We can argue in two ways against a double RGBBs due solely to differential reddening. First, if the magnitude difference between the two RGBBs in the \cnw\ is due to differential reddening, the spread in the interstellar reddening becomes $\Delta$\ebv\ $\approx$ \dvbump/3.1 $\approx$ 0.053 mag, which becomes $\Delta$\eby\ $\approx$ 0.04 mag, assuming \eby\ = 0.74$\times$\ebv. 
In other words, the intrinsic $(b-y)$ color (i.e., the reddening corrected color) of the lower part of RGBB at $(b-y) \approx$ 0.55 and $V \approx$ 15.6 mag in Figure~\ref{fig:bumpiso} should be 0.04 mag bluer at $V$ = 15.44 mag, i.e., $(b-y)_0$ = 0.51 mag. 
If so, the width of the dereddened \cnw\ RGB population becomes larger, $\Delta (b-y) \gtrsim$ 0.03 mag, than the reddened RGB width, $\Delta (b-y) \approx$ 0.02 mag. 

In panel (a) of Figure~\ref{fig:bumpiso} we show the \hkjwl\ versus $V$ CMD. The interstellar reddening of the \hkjwl\ index
is given by \ehk\ = $-$0.16$\times$\eby\ = $-$0.12$\times$\ebv\ \citep[see Supplementary Information of][and references therein]{lee09a}. If the differential reddening is responsible for the double RGBBs, then the dereddened \hkjwl\ color of the faint \cnw\ RGBB, 0.70, becomes about 0.05 mag redder, making the width of \hkjwl\ for the \cnw\ stars too broad for being a simple stellar population. Therefore, it is most likely that other effects are responsible for the double RGBBs in the \cnw\ population, as discussed below.

\subsection{Monte Carlo Simulations for the \cnw\ Double RGBBs}\label{ss:mc}
In order to understand the origin of the double RGBBs with a slightly broader RGB width in the \cnw\ population, we performed Monte Carlo simulations by constructing evolutionary population synthesis (EPS) models similar to those of our previous studies \citep[e.g.,][]{lc99b,lee15,lee19b}.

We have shown that at a fixed age, the RGBB magnitude is sensitively dependent both on the helium abundance and on the metallicity. In Figure~\ref{fig:bumpiso}  panels c-d, we present comparisons of model isochrones of \citet{valcarce12} to our observations.
Note that the $V$ magnitude levels of the RGBB of the model isochrones by \citet{valcarce12} are about 0.40 mag brighter than our observations. 
Recall that our RGBB $V$ magnitude for M3 is in excellent agreement with that of \citet{ferraro97}. In Figure~\ref{fig:bumpiso}, we add additional 0.40 mag to the model isochrones to match the location of the RGBB and the figure suggests that both a helium enhancement and a metallicity spread could explain the double RGBBs in the \cnw\ population. Using model isochrones by \citet{valcarce12}, we constructed two sets of EPS models: (1) an EPS model with \dy\ = 0.04 and $\Delta$[Fe/H] = $-$0.05 dex to reproduce the bright RGBB (Sim1, hereafter); and (2) an EPS model with $\Delta$[Fe/H] = $-$0.15 dex without the helium enhancement to reproduce the bright RGBB (Sim2, hereafter). Finally, we convolved our observational uncertainties as shown in Figure~\ref{fig:cmd} to generate synthetic CMDs. We performed 1,000 simulations for both sets to avoid  artifacts arisen from small sample sizes.

In our simulations, we adopt a fixed age for both Sim1 and Sim2, 12.5 Gyr. At a fix metallicity and helium abundance, the RGBB luminosity decreases with age and, therefore, the magnitude difference between the two RGBBs can be slightly tweaked with ages between the two presumed stellar populations. However, the previous \hst\ photometry suggested that M3 does not appear to show any perceptible age spread, 12.5 $\pm$ 0.50 Gyr \citep{dotter10}.

\begin{figure}
\epsscale{1.15}
\figurenum{22}
\plotone{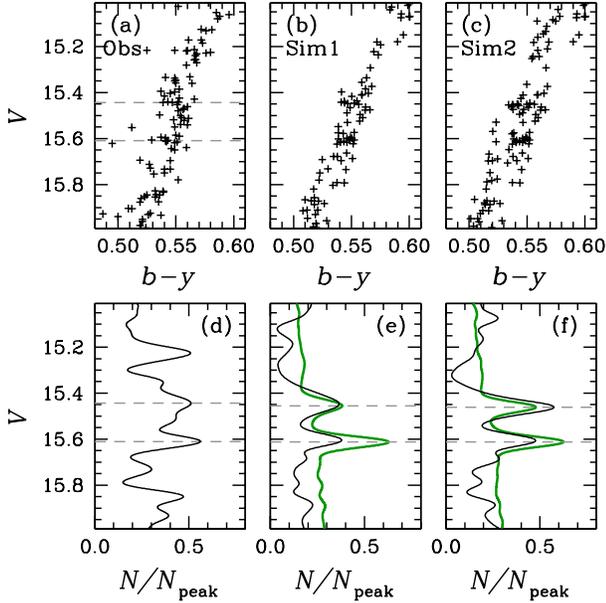}
\caption{
(a) CMD of the M3 \cnw\ RGB stars around the RGBB. The gray dashed lines indicate the double RGB peaks. 
(b) Synthetic CMD for Sim1, \dy\ = 0.04 and $\Delta$[Fe/H] = 0.05 dex.
(c) Synthetic CMD for Sim2, $\Delta$[Fe/H] = 0.15 dex. 
(d) Differential LF of (a).
(e) The black solid line shows the differential LF of (b), while the green solid line and the gray dashed lines represent the mean differential LF of 1000 trials and the $V$ magnitude of the double RGBBs.
(f) Same as (e), but for (c).
}\label{fig:eps}
\end{figure}

\begin{deluxetable}{ccc}
\tablenum{8}
\tablecaption{The $V$ magnitudes of RGBBs} \label{tab:rgbb}
\tablewidth{0pc}
\tablehead{\colhead{} &\colhead{bright RGBB}&\colhead{Faint RGBB}}
\startdata
Observations & 15.444 & 15.610 \\
Sim1 & 15.455 & 15.612 \\
Sim2 & 15.461 & 15.612 \\
\enddata 
\end{deluxetable}

\subsection{Helium Abundance: Sim1}\label{ss:rgbb_he}
In Figure~\ref{fig:bumpiso}(c), we show model isochrones for Sim1: [Fe/H] = $-$1.45 dex, Y = 0.25, and [Fe/H] = $-$1.50 dex, Y = 0.29 \citep{valcarce12}. Note that a small metallicity spread is required to reproduce the observed RGB width.
Our input models can explain the $V$ magnitudes of the double RGBBs satisfactorily.
But the RGB width between the two isochrones is slightly narrower than our observations.
To reproduce the RGB width of the isochrone for [Fe/H] = $-$1.50 dex, Y = 0.29 without metallicity and age spreads, a model with Y $\approx$ 0.33 is required as shown in the figure. However, its RGBB $V$ magnitude is about 0.08 mag brighter than that of the bright RGBB in our observations. An increased age by 1.0 -- 1.5 Gyr can reproduced the observed bright RGBB magnitude, but it is against the very narrow or no age spread of M3 from the \hst\ photometry \citep{dotter10}.

In Figure~\ref{fig:eps}(b) and Table~\ref{tab:rgbb}, we show the results from our Monte Carlo simulations. In the figure, we show a synthetic CMD returned from one particular simulation and its differential LF. As mentioned earlier, Sim1 shows a slightly narrower RGB width than our observations. We also show the average LF from 1,000 simulations, showing two RGBBs whose $V$ magnitudes are consistent with our observations.

In terms of the $V$ magnitudes of the double RGBBs and the \cnw\ RGB width, the helium enhanced model can reproduce our observations.
However, this large amount of helium enhancement in the \cnw\ population is hard to reconcile with the M3 HB morphology \citep[see also][]{tailo19}.
Also, assuming the metallicity is a chronometer, it is difficult to understand the chemical evolution from stars having low metallicity with high helium abundance ([Fe/H] = $-$1.50, Y = 0.29) to stars having slightly high metallicity but significantly low helium abundance ([Fe/H] = $-$1.45, Y = 0.25).
The helium abundance can be slightly decreased with dilution of pristine gas but the metallicity should be decreased, too. 
Without invoking a metallicity spread, it requires a large amount of helium enhancement, \dy\ = 0.08, and a large age spread, more than 1 Gyr, which are against the HB morphology \citep{tailo19} and the undetectably small age spread of M3\citep{dotter10}.
Therefore variations in the helium abundance do not appear to be responsible for the double RGBBs of the \cnw\ population.

\begin{figure}
\epsscale{1.2}
\figurenum{23}
\plotone{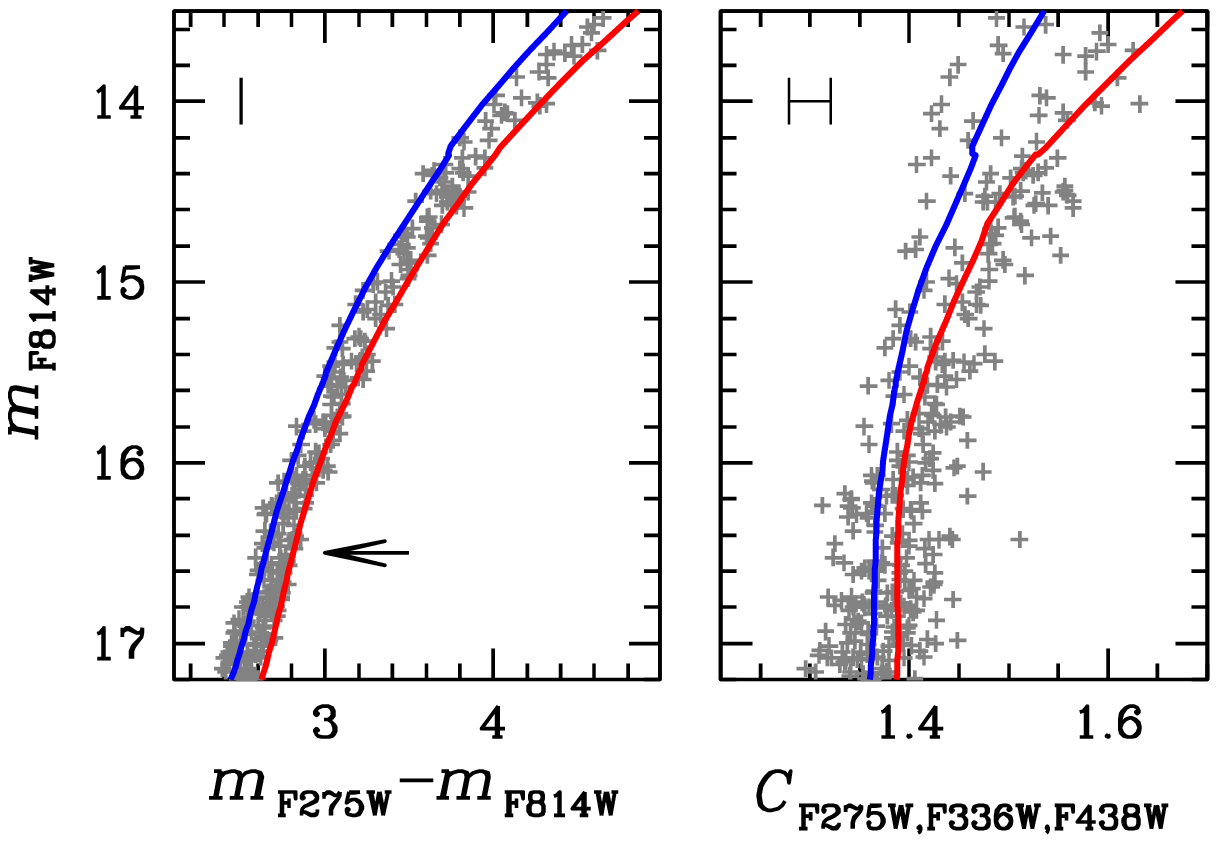}
\caption{
\hst\ pseudo-color CMDs for the M3 FG RGB stars using photometry by \citet{uvlegacy}.
The blue and red solid lines are for the Dartmouth isochrone with [Fe/H] = $-$1.60 and $-$1.45 dex, respectively. 
The arrow in the left panel indicates the $m_{\rm F814}$ magnitude where the \wfg\ is measured, 2 mag brighter than the main sequence turn-off. 
The thin error bars indicate the contribution of the spread in nitrogen abundance by $\Delta$[N/Fe] = 0.1 dex.
}\label{fig:hst2grid}
\end{figure}

\subsection{Metallicity: Sim2}\label{ss:rgbb_metal}
The M3 \cnw\ population has a metallicity spread of $\approx$ 0.07 -- 0.10 dex from high-resolution spectroscopy \citep{sneden04,apogee}, and we have derived a bimodal metallicity distribution with $\Delta$[Fe/H] $\approx$ 0.15 dex from \hkjwl\ photometry.
In Figure~\ref{fig:bumpiso}(c), we show the model isochrones  for Sim2: [Fe/H] = $-$1.60 and $-$1.45 dex with a fixed helium abundance, Y = 0.25. The figure shows that our input models can explain the double RGBBs and the RGB width reasonably well.
In the figure, we also show the model isochrone with a slight helium enhancement by $\Delta$Y = 0.01 at the metal-rich regime. 
It does not disagree with our observations but suggests that helium enhancement may not be the essential factor for
understanding the double RGBBs in M3.

In Figure~\ref{fig:eps}(c) and Table~\ref{tab:rgbb}, we show the results of our Monte Carlo simulations, indicating that both the $V$ magnitudes of the double RGBBs and the RGB width are consistent with our observations.
Therefore, we suggest that the discrete bimodal metallicity 
distribution must be the main reason for
the extended and tilted RGBB and the width of the RGB of the \cnw\ population.

Finally, we also compare the \hst\ photometry by \citet{uvlegacy} to model isochrones by \citet{dartmouth}, since \citet{valcarce12} does not provide magnitudes for the passbands required to construct the chromosome map.
In Figure~\ref{fig:hst2grid}, we show the Dartmouth model isochrones with [Fe/H] = $-$1.60 and $-$1.45 along with the M3 FG RGB stars using the results of \citet{uvlegacy}. Both model isochrones have Y = 0.25 and [$\alpha$/Fe] = 0.25 dex and
the distance modulus of 15.07 mag and the offset value of $-$0.10 mag in the ($m_{\rm F275W} - m_{\rm F814W}$) color were assumed.
\citet{milone17} measured the RGB width of the ($m_{\rm F275W} - m_{\rm F814W}$) color at 2 mag brighter than the main-sequence turn-off, obtaining \wfg\ = 0.244 $\pm$ 0.014 for the M3 FG population. As shown in the figure, model isochrones with different metallicity can nicely explain the M3 FG's ($m_{\rm F275W} - m_{\rm F814W}$) and [($m_{\rm F275W} - m_{\rm F336W})-(m_{\rm F336W} - m_{\rm F438W})$)] widths without invoking helium enhancement. We also calculate the contribution of the nitrogen abundance enhancement to the \wfg\ and our result confirms that nitrogen abundance does not affect the \wfg, as \citet[][and references therein]{milone17} suggested. On the other hand, the [($m_{\rm F275W} - m_{\rm F336W})-(m_{\rm F336W} - m_{\rm F438W})$)] color is sensitively dependent on nitrogen abundance since the F336W passband contains the NH band at \nhwave.

\medskip
We conclude that the M3 \cnw\ population has its double RGBBs because of the bimodal metallicity distribution.

\begin{figure}
\epsscale{1.2}
\figurenum{24}
\plotone{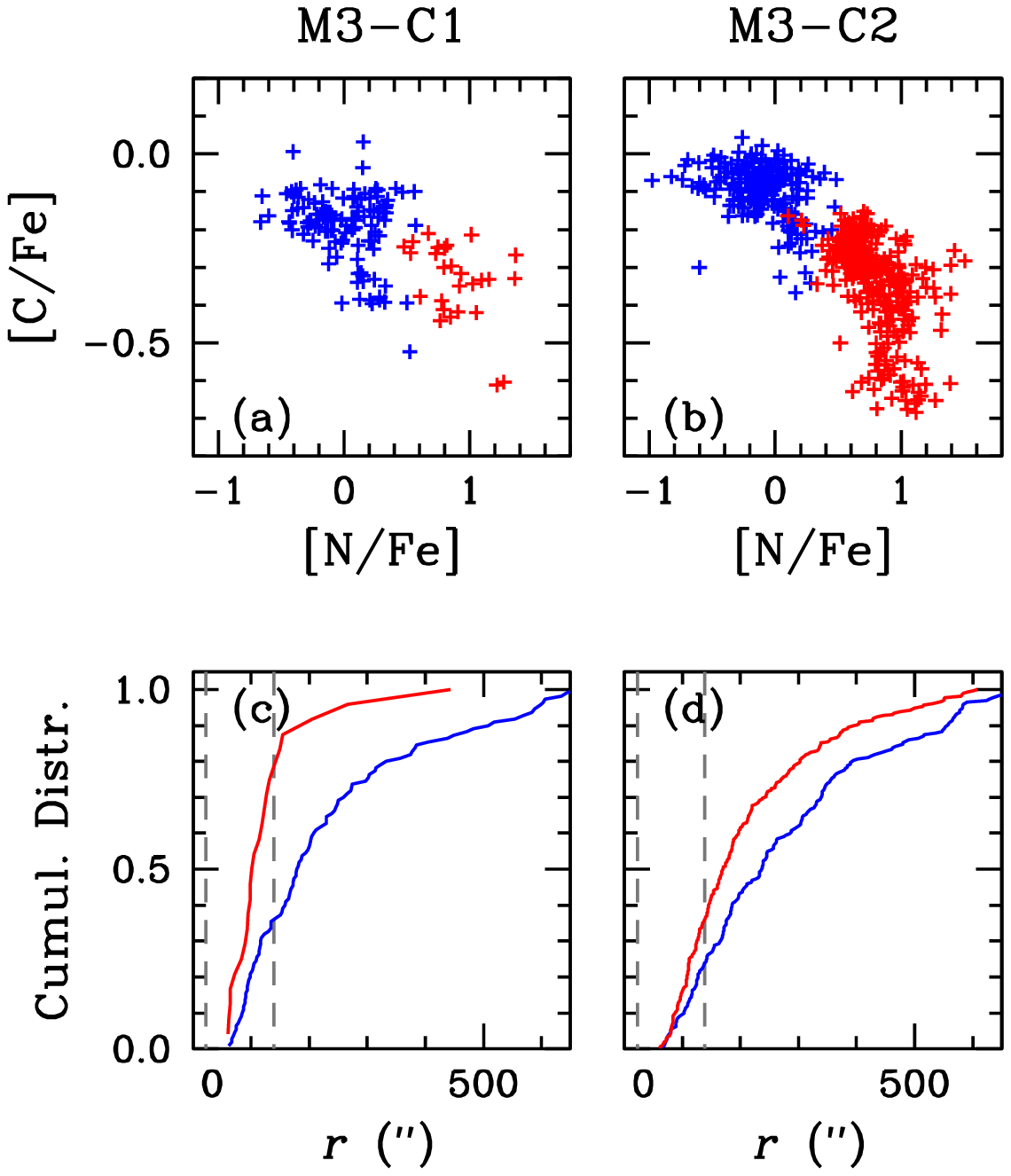}
\caption{
(a) [C/Fe] versus [N/Fe] distribution for the C1. The blue and red plus signs denote the C1:\cnw\ (i.e., \cnw:SP1) and C1:\cns\ (i.e., \cns:SP1), respectively.
(b) [C/Fe] versus [N/Fe] distribution for the C2. The blue and red plus signs denote the C2:\cnw\ (i.e., \cnw:SP2) and C2:\cns\ (i.e., \cns:SP2), respectively.
(c) The CRDs of the C1:\cnw\ (blue) and C1:\cns\ (red). 
The vertical gray dashed lines denote the core and half-light radii of M3.
(d) The CRDs of the C2:\cnw\ (blue) and C2:\cns\ (red). 
}\label{fig:2gc}
\end{figure}

\begin{figure}
\epsscale{1.15}
\figurenum{25}
\plotone{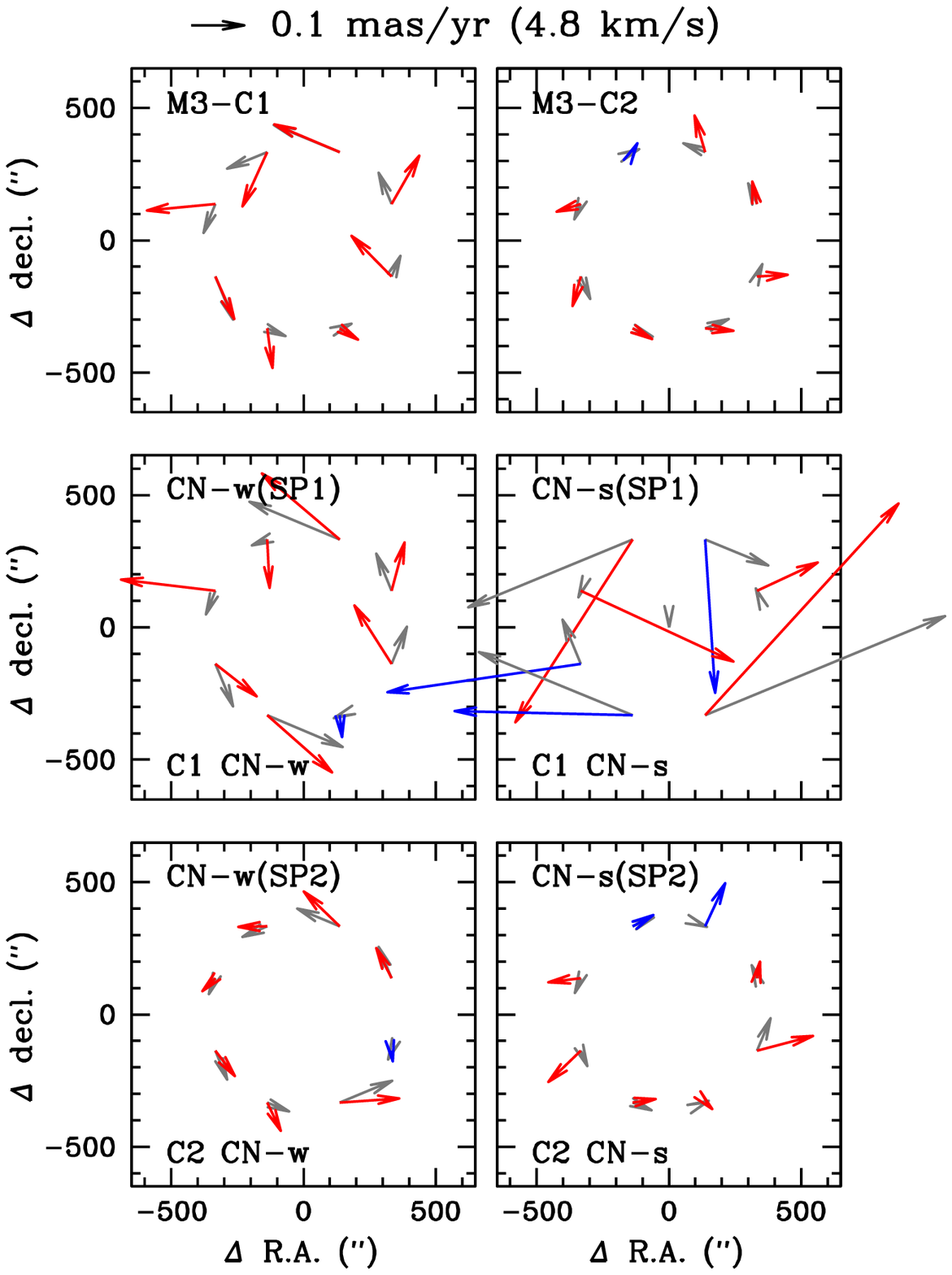}
\caption{
Mean proper motion vectors of eight slices in the radial zone of 1\arcmin\ $\leq$ $r$ $\leq$ 10\arcmin. The red arrows denote a counterclockwise rotation (E $\rightarrow$ N $\rightarrow$ W $\rightarrow$ S $\rightarrow$ E), while the blue arrow denote a clockwise rotation at a given position vector. The gray arrows represent the rotating component assuming a circular rotation. An arrow shown in above indicates a proper motion of 0.1 mas/yr or 4.8 \kms\ assuming 10.2 kpc for the distance to M3.
(Top panels) M3-C1 and C2.
(Middle Panels) C1:\cnw\ (i.e. \cnw:SP1) and C1:\cns\ (i.e. \cns:SP1)
(Bottom Panels) C2:\cnw\ (i.e. \cnw:SP2) and C2:\cns\ (i.e. \cns:SP2)
}\label{fig:2gcpm}
\end{figure}

\begin{figure}
\epsscale{1.15}
\figurenum{26}
\plotone{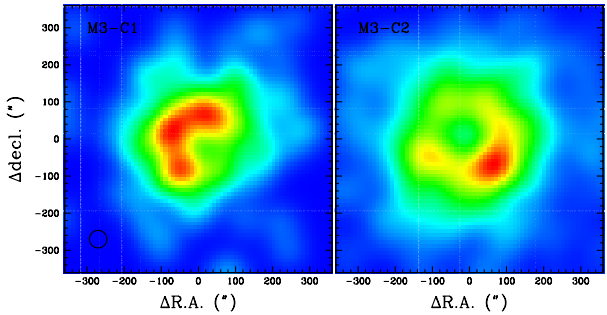}
\caption{
Smoothed density maps of the C1 (left) and C2 (right) RGB populations. The FWHM of our Gaussian kernel are shown with a circle. Note that the absolute peak values are different.
}\label{fig:2gcmap}
\end{figure}

\section{M3 as the merger remnant of two Globular Clusters}\label{s:merger}

\subsection{Discrete bimodal [Fe/H] distributions and a sequential formation scenario}
As we presented in Figure~\ref{fig:feh_hk} and Table~\ref{tab:subpop},  both the \cnw\ and \cns\ populations in M3 show discrete bimodal \fehhk\ distributions, $-$1.58 $\pm$ 0.05 (\cnw:SP1) and $-$1.44 $\pm$ 0.05 (\cnw:SP2) with the sub-populational number ratio of 36:64 ($\pm$4) for the \cnw\ population and  $-$1.59 $\pm$ 0.05 (\cns:SP1) and $-$1.47 $\pm$ 0.03 (\cns:SP2) with the sub-populational number ratio of 17:83 ($\pm$2) for the \cns\ population. In each population, the mean metallicities of the two sub-populations do not overlap within a 1$\sigma$ level.

Discrete bimodal metallicity distributions in both populations cannot be produced by incomplete mixing in their proto-cluster clouds and they require episodic star formations. Furthermore, we find it impossible to explain our results of the elemental abundances in M3 with a sequential star formation scenario \citep[see also,][]{lee15}. Assuming our \cnw\ population is the FG and the \cns\ is the SG, the sequential formation scenario for M3 may be describe in the following steps: (1) The formation of the \cnw:SP1 with high [C/Fe] and low [N/Fe] abundances; (2) Type II supernovae (SNe II) explosion which enriched metallicity of the \cnw:SP2 without changing [C/Fe] and [N/Fe] abundances; (3) The formation of the \cns:SP1 out of the gas experienced CN-cycle hydrogen burning and an unknown process that decreases metallicity from [Fe/H] = $-$1.45 to $-$1.60; (4) SNe II explosion which enriched metallicity of the \cns:SP2 without changing [C/Fe] and [N/Fe] abundances. 

Alternative scenarios with some variations could also be possible but we expect at least four difficulties in any sequential formation scenarios: (1) The whole formation time scale should be no larger than 0.5 Gyr, as the \hst\ photometry of \citet{dotter10} suggested. (2) It requires some unknown processes that reduce metallicity before the formation of the \cns\ population. (3) To retain ejecta from energetic SNe II explosions requires much massive system for M3 in the past, most probably the relic of the more massive primeval dwarf galaxy \citep[e.g.,][]{lee09a,lee15}.
(4) If the metal-rich components were the SG of M3, they tend to be more centrally concentrated \citep[e.g., see][for $\omega$ Cen]{bellini09}, which is against our results as discussed below.

\subsection{A merger scenario}
A more simple and plausible explanation can be found in a merger of two GCs, {\it \`{a} la} M22. 
We have shown that the peak metallicity values and metallicity dispersions coincide both in the \cnw\ and \cns, \fehhk\ $\approx$ $-$1.60 and $-$1.45 with the metal rich sub-populations being the major components in both populations.
If we divide individual sub-populations with metallicity, the formation of M3 can be depicted in a simple way: the merger of two GCs, most probably, in a dwarf galaxy environment where the GC merger rates are significantly higher than those of our Galaxy, since the relative velocity of the two GCs in a dwarf galaxy is smaller than the velocity dispersions \citep[e.g.,][]{thurl02,lee09a,bekkiyong12,lee15,gavagnin16}.
Let us suppose that individual sub-populations can be re-arranged as follows: (1) The C1 system with [Fe/H] = $-$1.60, which is composed of the \cnw:SP1 and the \cns:SP1. In the C1, the \cnw:SP1 corresponds to the C1:\cnw\ and the \cns:SP1 is to the C1:\cns, with the number ratio of \nrgb\ = 81:19 ($\pm$5). (2) The C2 system with [Fe/H] = $-$1.45, which is composed of the \cnw:SP1 and the \cns:SP1. 
Similar to C1, in C2 the \cnw:SP2 corresponds to the C2:\cnw\ and the \cns:SP2 is to the C2:\cns, with the number ratio of 43:57 ($\pm$4).
We caution that our sub-populational number ratios for the C1 and C2 could be different from their true values, since we relied on the stars located at $r \geq 1\arcsec$ and $\sigma(hk) \leq$ 0.01 mag. The latter constraint also prefers stars in the outer region, where the degree of crowdedness is less severe, resulting in small measurement errors.

Recently, \citet{milone20} studied GC MPs both in our Galaxy and Magellanic Clouds (MC), finding that the FG fractions of MC clusters are significantly higher for their masses, from about 50\% upto 80\%, than Galactic GCs, most likely reflecting environmental effect.
The current mass of M3 from \citet{baumgardt18} is $\approx$ $3.9\times10^5$\msun. If we adopt the number ratio between the two system, $n$(C1):$n$(C2) = 23:77 ($\pm$2), then the total masses for the C1 and C2 become $\approx$ $0.9\times10^5$\msun\ and  $3.0\times10^5$\msun, respectively. If our \cnw\ is the same as the FG of \citet{milone20}, the fractions of the FG of the C1 and C2, 81\% and 43\%, respectively, appear to follow the trend of the MC GCs (see their Figures 7 and 12).

In Figure~\ref{fig:2gc}(a--b), we show the [C/Fe] versus [N/Fe] anticorrelations for the C1 and C2. The C1 shows a less extended C-N anticorrelation than the C2, which may be consistent with the fact that less massive GC systems show less extended Na-O anticorrelations \citep[e.g.,][]{carretta09}.

Being a less massive system, the internal helium enhancement of C1 must have been smaller than that of C2.
At the current mass of C1, the results of \citet{milone18} suggest that the helium enhancement of the C1:\cns\ (= \cns:SP1) sub-population is almost nil, which may explain the absence of the double RGBBs in our \cns\ population as shown in Figure~\ref{fig:rgbb}. Without helium enhancement, the RGBB $V$ magnitudes of the \cnw:SP1 and \cns:SP1 would be the same, consistent with our observations, $V$ = 15.444 and 15.446 mag. In addition, the fraction of the metal-poor component of the \cns\ population, the C1:\cns\ (= \cns:SP1), appears to be too low to produce conspicuous double RGBBs in the \cns\ population.

Figure~\ref{fig:2gc}(c--d) show cumulative radial distributions of individual sub-populations in the C1 and C2. In both systems, the \cns\ sub-populations are more centrally concentrated than the \cnw, consistent with those of other Galactic GCs. In Table~\ref{tab:2gcks}, we show $p$-values returned from the K-S tests for CRDs of the C1 and C2. The \cnw:SP1 (i.e., the C1:\cnw) and \cns:SP2 (the C2:\cns) shows a high $p$-value. It is not clear whether they are physically linked together or their similar CRDs are pure coincidence. The $p$-values for other combinations are very low and their CRDs are probably not related to each other.

We also examine the proper motion of C1 and C2, presenting our results in Figure~\ref{fig:2gcpm}. C1 does not appear to show any aligned projected rotation and it appears to be a random motion dominated system, while C2 shows a well aligned projected circular motion. Our results strongly suggested that C1 and C2 have different internal motions and they are not kinematically homogenized.

Finally, we examined the spatial distributions and we show the smoothed density map of the C1 and C2 in Figure~\ref{fig:2gcmap}. The figure clearly shows that the centers of the C1 and C2 are most likely different. We tried to derived their centers but, unfortunately, the coordinates of each center without stars in the central part of M3 are significantly different depending on methods that we adopted in our previous study \citep{lee15} and failed to yield reliable coordinates. 
Note that Figure~\ref{fig:2gcmap} can be somewhat misleading. Due to lack of stars in $r < 1\arcmin$, asymmetries both in the C1 and C2 are exaggerated during our smoothing process. In the future, it would be very desirable to investigate the central part of M3 using our photometric system, which will reveal the structural properties of the C1 and C2.

\begin{deluxetable}{cc}
\tablenum{9}
\tablecaption{ 
$p$-values returned from the K-S Tests for CRDs of M3 C1 and C2} \label{tab:2gcks}
\tablewidth{0pc}
\tablehead{\colhead{} &\colhead{$p$-value}}
\startdata
\cnw(SP1) vs. \cnw(SP2) & 0.005 \\
\cns(SP1) vs. \cns(SP2) & 2.3$\times10^{-7}$ \\
\cnw(SP1) vs. \cns(SP1) & 6.8$\times10^{-5}$ \\
\cnw(SP1) vs. \cns(SP2) & 0.852 \\
\cnw(SP2) vs. \cns(SP1) & 2.4$\times10^{-8}$ \\
\cnw(SP2) vs. \cns(SP2) & 2.4$\times10^{-5}$ \\
& \\
\cnw\ vs. \cnw(SP1) & 0.548 \\
\cnw\ vs. \cnw(SP2) & 0.009 \\
\cnw\ vs. \cns(SP1) & 7.8$\times10^{-6}$ \\
\cnw\ vs. \cns(SP2) & 0.010 \\
& \\
\cns\ vs. \cnw(SP1) & 0.004 \\
\cns\ vs. \cnw(SP2) & 1.2$\times10^{-14}$ \\
\cns\ vs. \cns(SP1) & 0.015 \\
\cns\ vs. \cns(SP2) & 0.001 \\
\enddata 
\end{deluxetable}

\section{Summary and Conclusion}\label{s:sum}

As part of our ongoing effort to investigate MPs in GCs based on homogeneous photometry, we developed a new filter, $JWL34$, that can measures the NH bands at \nhwave\ of cool stars and a new photometric index, \nhjwl, a photometric measure of nitrogen abundance.
Our new photometric system combined with the robust and self-consistent 
theoretical fine model grids for various parameter sets using synthetic spectra allows us to measure key elements in stellar populations, [Fe/H], [C/Fe] and [N/Fe], even in the extremely crowded fields, such as the central part of GCs, where the traditional spectroscopic observations cannot be performed.
Unlike other broad-band photometric systems, such as those used in the \hst\ photometry, our photometric indices, \hkjwl, \nhjwl\ and \chjwl, suffer less degree of degeneracy of various elemental abundances and they can provide straightforward solutions. 
Our approach appears to be a good pathway to study GC MPs even with small aperture ground-based telescopes. It is a powerful tool to secure large sample sizes that permit us to reveal substructures in M3 that cannot be seen with the more limited samples available in the past.

In this paper, we investigated the MPs of the prototypical normal GC M3. We derived the populational number ratio based on the merged data (i.e., our photometry in the outer part and the \hst\ photometry in the central part of M3), obtaining \nrgb\ = 38:62 ($\pm$2), which is different from our previous estimate, 48:52 ($\pm3$), due to incomplete detection of stars in the central part of M3 from our ground-based observations with a small aperture telescope. Our revised populational number is still different from that of \citet{milone17}, who obtained the FG fraction of 0.305 $\pm$ 0.014. It is an apparently simple number counting task, but our results for the populational number ratio vividly show the importance of securing an unbiased sample from the large sampling area and complete detection of sources, which always has been a core difficulty in astronomy.

Comparisons of our color indices with those of the \hst\ \citet{milone17} showed that the \cnw\ and \cns\ have different correlations with the \hst\ color indices, mainly due to different elemental abundances between the two populations and the different degree of sensitivity of individual passbands with significantly different bandwidths on elemental abundances.

The derived \fehhk\ from our \hkjwl\ showed bimodal metallicity distributions for both the \cnw\ and \cns\ populations, with the peak metallicities of \fehhk\ = $-$1.60 and $-$1.45, which we believed to be the mean metallicities of the two GC populations in M3.

Assuming the [O/Fe] from \citet{apogee}, we obtained [C/Fe] = $-$0.10 dex, [N/Fe] = 0.10 dex for the less evolved \cnw\ fRGB, while [C/Fe] = $-$0.30 dex, [N/Fe] = 0.80 dex for the less evolved \cns\ fRGB. Our carbon and nitrogen abundances from the CH and NH bands are in reasonable agreement with those of other low resolution spectroscopy studies.
However, our carbon and nitrogen abundances from the CN band at \cnwave\ are about 0.2 dex higher than those from the CH and NH bands. Our calculations showed that  high input oxygen abundances that we adopted from \citet{apogee} are most likely responsible for our high carbon and nitrogen abundances from the CN band.

Evidence from many studies, including our previous papers, have shown that the chemical evolution from the \cnw\ population to the \cns\ population is not continuous.  
For example, the plot of the [C/Fe] versus [N/Fe] clearly shows two separate relations.
The carbon abundance does not appear to be strongly correlated with nitrogen abundance in the \cnw\ population, indicating that the \cnw\ population formed out of interstellar media that experienced little or no CN-cycle hydrogen burning. On the other hand the \cns\ population shows a C-N anticorrelation, likely due to a natural consequence of the CN-cycle occurred in the previous generation. 

We showed that the \cnw\ and \cns\ populations are also different in their structural and kinematical properties. We confirmed our previous result that the M3 \cns\ population is more centrally concentrated than the \cnw\
is. On the other hand, the \cnw\ shows a more elongated morphology with a well-aligned circular motion.

As an extension of the metal-rich RHB studies by others \citep[e.g.,][]{norris82,smith89}, we showed that the \cnjwl\ index can be used to study the RHB stellar population of the intermediate metallicity GCs.
We derived the populational number ratio for the M3 RHB, finding \nrgb\ = 94:6 ($\pm$3), which is significantly different from that of the RGB but can be naturally explained in the evolution of the metal-poor low-mass stars. The helium normal \cnw\ M3 RGB stars evolve into the RHB or RRLs, while the helium enhanced \cns\ RGB stars evolve into the BHBs or RRLs, which is also supported by the synthetic HB model simulations of \citet{tailo19}.
Our K-S tests for the M3 RGB and HB showed that the RHB is the \cnw\ population, while the BHB is the \cns. The RRL is a mixed population.

We discussed the existence of the discrete double RGBBs in the \cnw\ population.  
\cite{lee19b} suggested that the \cnw\ population exhibits an extended and titled RGBB. However, our current study clearly shows that the \cnw\ RGBB can be well described with the discrete double RGBBs, which are consistent with the \hst\ photometry by \citet{uvlegacy}.

To understand the origin of the discrete double RGBBs, we performed Monte Carlo simulations using our EPS models, finding that M3's bimodal metallicity distribution, [Fe/H] = $-$1.60 and $-$1.45, can nicely reproduce our observations. The enhanced helium model cannot reproduce our observations without some metallicity span ($\Delta$[Fe/H] $\geq$ 0.05 dex) or a large age spread  ($\Delta$Age $\geq$ 1 Gyr), which is also supported by the results of \citet{tailo19} that helium enhancement cannot fully explain the M3 HB morphology.
At the same time, the bimodal metallicity distribution can explain the large M3 FG \wfg\ range that has puzzled others \citep{lardo18,tailo19}.

Finally, we discussed the new formation history of M3. From our new discovery of the bimodal metallicity distributions for both the \cnw\ and \cns\ populations, which is supported by the discrete double RGBBs and the large \wfg\ range, we proposed that M3 can be well described by the two GCs, namely, the C1 (23\% of our sample with $\langle$[Fe/H]$\rangle$ $\approx$ $-$1.60) and the C2 (77\% of our sample with $\langle$[Fe/H]$\rangle$ $\approx$ $-$1.45). In both systems, the \cns\ components are more centrally concentrated, which is the common property of Galactic GCs with MPs. At the same time, the \cnw\ and \cns\ components in each system share the common kinematical properties: The C1 appears to be a random motion dominated system, while the C2 shows a more well-aligned rotation. We derived the sub-populational number ratios, finding that \nrgb\ = 81:19 ($\pm$5) for the C1 and 43:57 ($\pm$4) for the C2. Given the current mass of M3, the sub-populational number ratios of the C1 and C2 compared to Galactic GCs are large for their individual masses, but they are in good agreement with those of GCs in Magellanic Clouds \citep[e.g., see][]{milone20}. Therefore, it is most likely that M3 is a merger remnant of two GCs, likely in a less massive environment, such as dwarf galaxies where the GC merger rates are higher than our Galaxy \citep{thurl02,bekkiyong12,lee15,bekki19}, and accreted to our Galaxy later in time, which is consistent with the origin of the {\it Helimi Stream} as proposed by \citet{koppelman19}.

Although it is somewhat speculative, a merger scenario could help to understand the neglected mysteries of the RRL richness of M3. For example, a merger of two GCs, such as M5 ($\approx$ 130 RRLs and [Fe/H] = $-$1.47) and IC 4499 ($\approx$ 100 RRLs and [Fe/H] = $-$1.53) can explain the RRL richness of M3 ( $\approx$ 240 RRLs and [Fe/H] = $-$1.50).

Due to limitations in spatial angular resolving power of small aperture ground-based telescope that we employed, we were not able to perform a reliable study in the central part of M3 ($r \leq$ 1\arcmin), where the main body of M3 resides.
The follow up study using our new photometric system and large aperture telescopes to obtain high precision photometry of the central part of M3 will shed more light on our new discovery presented in this paper.

\acknowledgements
We thank an anonymous referee for a careful review of the paper and many helpful suggestions.
J.-W.L.\ acknowledges financial support from the Basic Science Research Programs (grant nos.\ 2016R1A2B4014741 and 2019R1A2C2086290) through the National Research Foundation of Korea (NRF) and from the faculty research fund of Sejong University in 2019.
C.S. acknowledges support from the U.S. National Science Foundation grant AST-1616040.

\appendix\label{s:ap}
The synthetic model grids used in our study depend on several parameters. Here, we investigate three input parameters ([O/Fe], \ciso\ and [Fe/H]) for synthetic model grid calculations that can affect our photometric [C/Fe], [N/Fe] and \fehhk. 
In addition, we also investigate the effect of carbon abundance on the \fehhk.  
Note that elemental abundance measurements from any low- to intermediate-resolution spectroscopy are also vulnerable to some uncertainties discussed here.

\section{Oxygen abundance}\label{s:ap:ofe}
Molecular CN and CH band strengths in RGB stars can be affected by not only the carbon and nitrogen abundances but also the oxygen abundance through the formation of the CO molecule in their atmospheres. In order to examine the influence of oxygen, we calculated individual color indices using the method described in \S\ref{s:abund} with different input oxygen abundances. For our calculations, we assumed [Fe/H] = $-$1.50 for both populations with different CNO abundances: [C/Fe] = $-$0.10, [N/Fe] = 0.10, and [O/Fe] = 0.50 and 0.40 for the \cnw\ population and [C/Fe] = $-$0.30, [N/Fe] = 0.80, and [O/Fe] = 0.15 and 0.05 for the \cns\ population (see Table~\ref{tab:CNabund}).

\begin{figure}
\epsscale{1.}
\figurenum{27}
\plotone{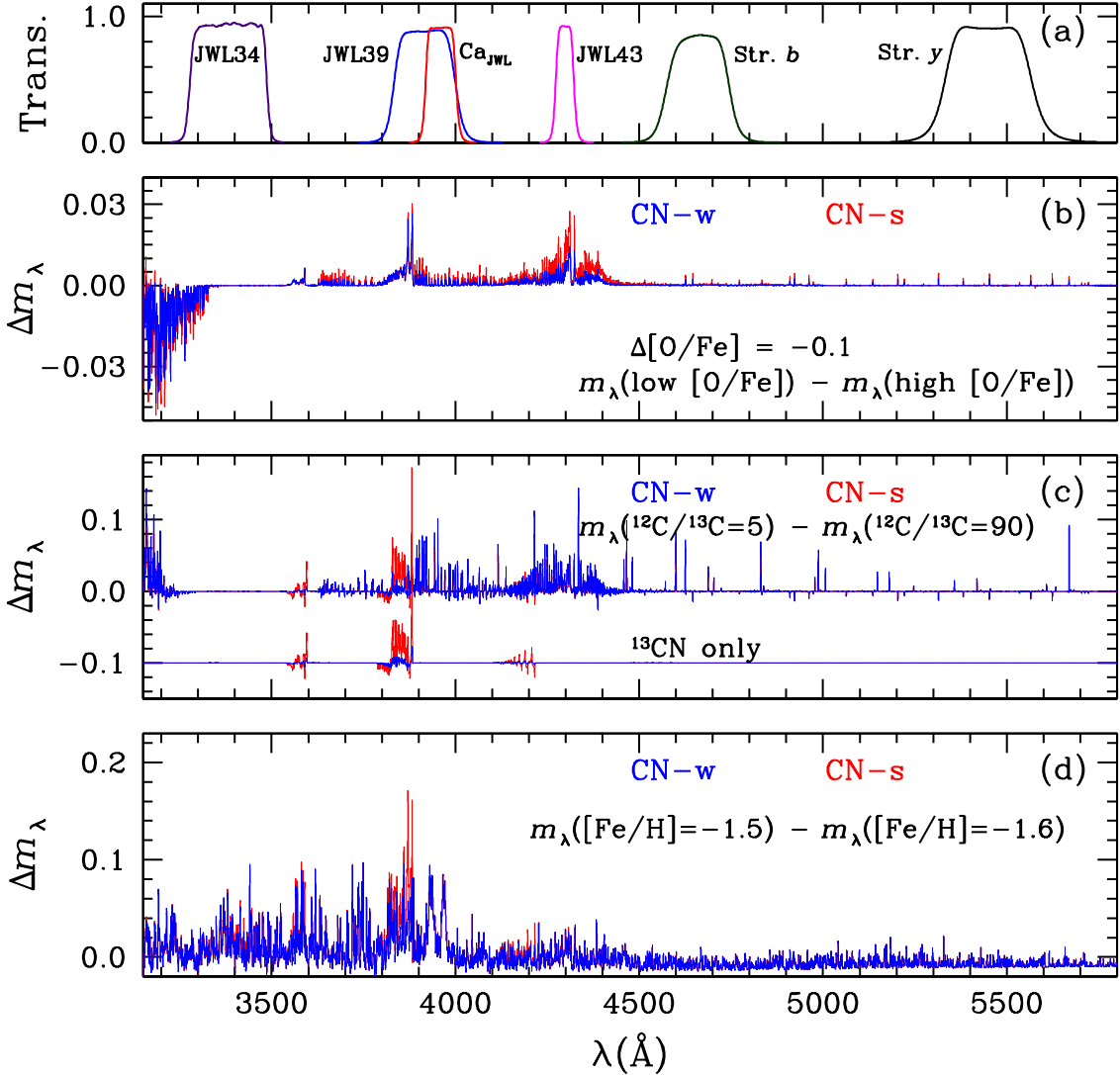}
\caption{
(a) Filter transmission functions.
(b) The difference in the monochromatic magnitude between the model atmosphere with $\Delta$[O/Fe] = $-$0.1. We show the results for $M_V$ = 0.0 mag. The blue and red solid lines denote the \cnw\ and \cns, respectively.
(c) Same as (b) but for \ciso\ = 5 and 90. The lower lines show differences due only to the CN contribution, with an offset of $-$0.1 mag.
(d) Same as (b) but for $\Delta$[Fe/H] = 0.1 dex.
}\label{fig:ap:comp_syn}
\end{figure}

In Figure~\ref{fig:ap:comp_syn}(a--b), we show filter transmission functions and differences in monochromatic magnitudes, in the sense of those of low [O/Fe] minus those of high [O/Fe]. The different oxygen abundance affects our color indices in two ways: 
(1) In the wavelength region shorter than $\approx$ $\lambda$3300\AA, the enhanced oxygen abundance strengthens the OH band features, part of which lie within the $JWL34$ bandwidth, as shown in Figure~\ref{fig:ap:comp_syn}(b). (2) It weakens the CN and CH band features since a considerable amount of carbon atoms are tied up in CO molecules.
On the other hand, the variation of the CN abundance due to varying oxygen abundance does not affect the NH populations at optical depths of the absorption line formation. Therefore, our \nhjwl\ is affected only by the OH band features, as we mentioned above.

\begin{figure}
\epsscale{1.}
\figurenum{28}
\plotone{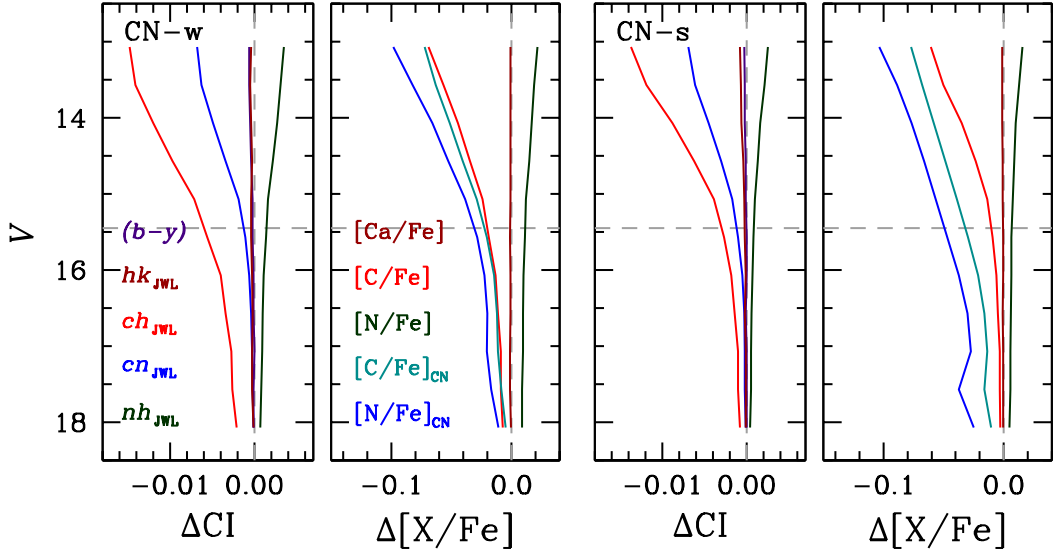}
\caption{
Differences in color indices and inferred elemental abundances due to different [O/Fe].
}\label{fig:ap:ofe}
\end{figure}

Figure~\ref{fig:ap:ofe} shows how oxygen abundance affects our color indices and photometric elemental abundances. As shown, oxygen abundance does not affect the $(b-y)$ and \hkjwl. On the other hand, our \nhjwl\ index is slightly affected by oxygen abundances, not by different NH population in the atmosphere but by OH contribution in the $JWL34$ bandwidth, as we mentioned above.
The  \cnjwl\ and \chjwl\ can be more seriously affected by oxygen abundances.
At a fixed carbon abundance, as oxygen abundance increases, the the  \cnjwl\ and \chjwl\ index values decrease due to decrease in the CN and CH populations.
As consequences, an underestimated oxygen abundance would result in higher photometric \cfecn\ from the \chjwl\ and \cnjwl\ indices and higher photometric \nfecn\ from the \cnjwl\ index. At bright RGB stars, $\Delta$[O/Fe] = 0.1 dex can result in $\Delta$[C/Fe] $\approx$ 0.06 -- 0.07 dex from \chjwl\ and \cnjwl, and  $\Delta$\nfecn\ $\approx$ 0.10 dex from \cnjwl\ at bright RGB regime for both populations. On the other hand, the oxygen abundance does not affect the [N/Fe] from \nhjwl, affecting no larger than 0.02 dex.

\section{\ciso\ ratios}\label{s:ap:ciso}
The carbon isotope ratio, \ciso, can also affect some of our color indices \citep[e.g., see][]{briley89,lee19c}. In GC RGB stars, as a result of the CN process at high temperature accompanied by a non-canonical thermohaline mixing \citep[e.g., see][]{charbonnel07}, the surface carbon $^{13}$C isotope abundance increases and the \ciso\ ratio decreases. 
Inhomogeneous \ciso\ ratios among different populations in a given GC also can be produced depending on the contribution of the CN processes in previous generation of stars, supported by the existence of inhomogeneous nitrogen abundances in GC stars. 
Unfortunately \ciso\ ratios in large-enough samples of GC stars are largely unknown.

In Figure~\ref{fig:ap:comp_syn}(c), we show the differences in monochromatic magnitude between synthetic spectra with an enhanced $^{13}$C abundance (\ciso\ = 5) and a solar $^{13}$C abundance (\ciso\ = 90). In that panel, we show two cases:
(1) those including contributions from both CN and CH molecules, and 
(2) just those from the CN molecule (C$_2$ transitions are undetectably weak). The contribution from $^{13}$CN clearly is much smaller than that from $^{13}$CH. Only in the CN \cnwave\ band does the enhanced $^{13}$C abundance make any spectroscopic difference (as shown by \citealt{briley89}).

Our JWL34 filter is almost free from $^{13}$CH contamination and, as a consequence, our [N/Fe] measurements from the \nhjwl\ index is hardly affected by the varying $^{13}$C abundance. The \nhjwl\ index is very weakly dependent on the \ciso\ ratios due to some weak CH band features in the \str\ $b$ and $y$ passbands.

There exist some $^{13}$CH features within the \cajwl\ band. However, the \hkjwl\ index is much more sensitively dependent on changes in metallicity with a fixed [Ca/Fe] than it is to $^{13}$CH. 
\fehhk\ is essentially independent from $^{13}$CH contamination.

\begin{figure}
\epsscale{1.}
\figurenum{29}
\plotone{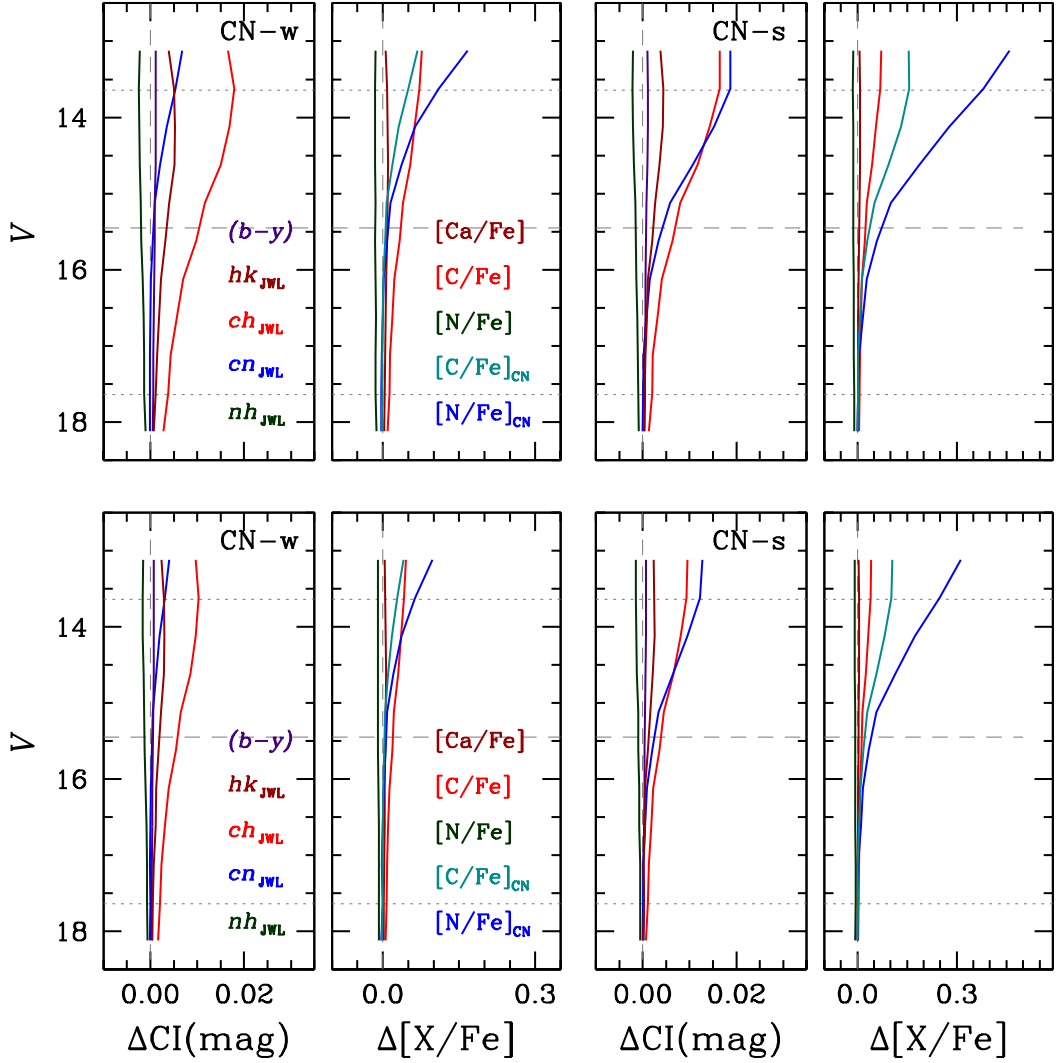}
\caption{
Differences in color indices and inferred elemental abundances due to different \ciso\ ratios. The upper panels are for \ciso\ = 5 and 90, and the lower panels are for \ciso\ = 5 and 10.
}\label{fig:ap:ciso}
\end{figure}

Figure~\ref{fig:ap:ciso} shows how \ciso\ ratios affect the color indices and elemental abundances inferred from color indices. In the figure, we show for the two cases: (1) \ciso\ = 5 and 90, and (2) \ciso\ = 5 and 10.  
The enhanced $^{13}$CN and $^{13}$CH abundances can make substantial differences in both the \nfecn\ and \cfecn\ values measured from the \cnjwl, in particular for bright RGB stars. It also affects the [C/Fe] from \chjwl.
As shown in the figure, the difference in the inferred \nfecn\ can be as large as 0.5 dex, in the sense that the enhanced $^{13}$CH abundance can make stars to be seen more [N/Fe]-rich.

\section{Metallicity}\label{s:ap:feh}
Metallicity also affects our color indices. In Figure~\ref{fig:ap:comp_syn}(d), we show monochromatic magnitude differences for $\Delta$[Fe/H] = 0.1 dex. Metallicity affects our color indices in two ways: (1) The continuum opacity from H$^-$ ion and Rayleigh scattering from neutral hydrogen depends on metallicity, since the formation of H$^-$ ion diminishes as metallicity decreases. (2) The overall absorption strengths increase with metallicity. The former effect can be minimized by using local continuum sidebands as practiced in low- to intermediate-resolution spectroscopy. However, inappropriate local continuum sideband assignment could bring a serious problem by setting incorrect continuum levels in the forest of heavily absorption lines \citep[e.g., see][for the case of the CH G-band]{lee19c}. Our \nhjwl\ index, for example, can be affected not only by individual absorption strengths but also by the shape of continuum since the baseline of our \nhjwl\ (the difference in the pivot wavelengths between the $JWL34$ and \str\ $y$) is as large as $\Delta\lambda\approx$ 2000\AA. 

\begin{figure}
\epsscale{1.}
\figurenum{30}
\plotone{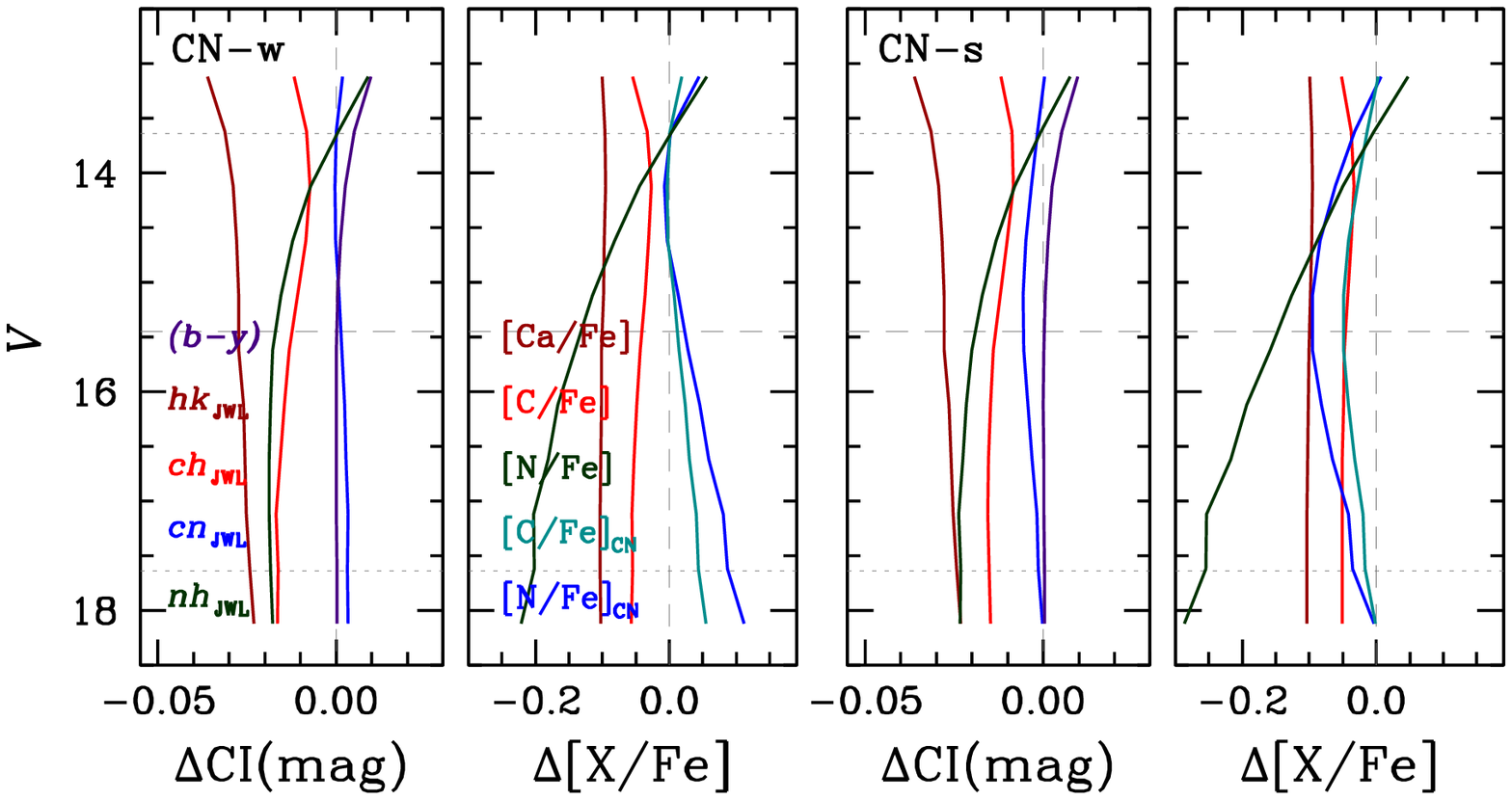}
\caption{
Differences in color indices and inferred elemental abundances due to different metallicity.
}\label{fig:ap:feh}
\end{figure}

In Figure~\ref{fig:ap:comp_syn}(c), we show the differences in monochromatic magnitude between synthetic spectra with different metallicities, where the slight difference in continuum opacity can be seen.
Figure~\ref{fig:ap:feh} shows the influence of metallicity on our color indices. As shown, [N/Fe] estimated from the \nhjwl\ is affected most, in particular in the faint RGB stars. In our simulations, $\Delta$[Fe/H] = 0.1 dex can result in $\Delta$[N/Fe] = 0.2 dex. 
Happily, the change in [Ca/Fe] the change in [Fe/H], confirming \hkjwl\ as a good metallicity index.

\section{Effect of Carbon Abundance on \hkjwl\ and \fehhk}\label{s:ap:c2ca}
\begin{figure}
\epsscale{1.}
\figurenum{31}
\plotone{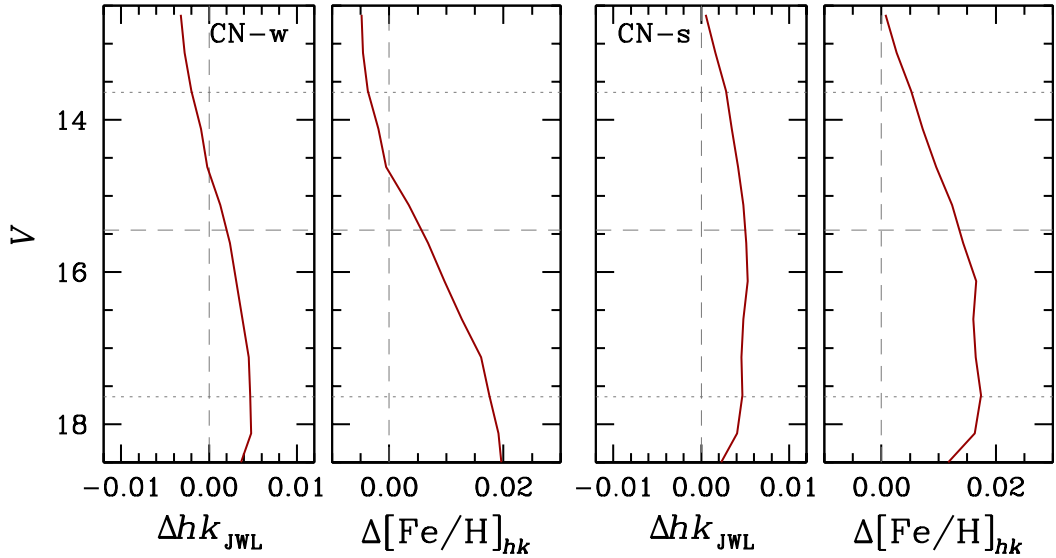}
\caption{
Differences in the \hkjwl\ and \fehhk\ due to the enhancement of carbon abundance by $\Delta$[C/Fe] = 0.2 dex.
}\label{fig:ap:CFe2CaFe}
\end{figure}

In our previous work \citep{lee19c}, we discussed that the passband of our \cajwl\ filter contains weak CH lines, which is also shown in Figure~\ref{fig:ap:comp_syn}(b--c).
In Figure~\ref{fig:ap:CFe2CaFe}, we show the effect of the carbon abundance enhancement by $\Delta$[C/Fe] = +0.2 dex on our \hkjwl\ and \fehhk. Our results strongly suggest that the spreads in carbon abundance in both populations, $\sigma$[C/Fe] $\leq$ 0.13 dex, are not responsible for the bimodal \fehhk\ distribution with the difference between the two peaks of $\Delta$\fehhk\ = 0.15 dex.

\end{document}